\documentclass[11pt,a4paper]{article}
\usepackage[russian,english]{babel}
\pdfoutput=1
\usepackage{cite}
\usepackage{amssymb,amsmath}
\usepackage{jheppub}
\usepackage[T1]{fontenc}
\usepackage{caption}
\usepackage{multirow}
\usepackage{tabularx}
\usepackage{makecell}
\usepackage{floatrow}
\usepackage{color}

\def\be{\begin{eqnarray}}
\def\ee{\end{eqnarray}}

\begin{document}
\begin{flushright}
INR-TH-2017-007
\end{flushright}

\title{\boldmath  Renormalization scheme  and gauge  (in)dependence 
of the generalized Crewther relation: what 
are the real grounds of the $\beta$-factorization property?}


\author[a,1]{A. V. Garkusha,\note{Affiliation from 2011 till 2015.}}
\author[b,c]{A. L. Kataev,}
\author[b,c]{V. S. Molokoedov}

\affiliation[a]{Higher School of Economics, Math Department,\\ Myasnitskaya street 20, Moscow, 101000, Russia}
\affiliation[b]{Institute for Nuclear
Research of the Russian Academy of Sciences,\\60th October Anniversary prospect 7a, Moscow, 117312, Russia}
\affiliation[c]{Moscow Institute of Physics and Technology,\\Institusky per. 9, Dolgoprudny, Moscow region, 141700, Russia}

\emailAdd{kataev@ms2.inr.ac.ru}
\emailAdd{viktor\_molokoedov@mail.ru}

\abstract{The problem of scheme and gauge dependence  of the factorization property of the renormalization group $\beta$-function
  in the $SU(N_c)$  QCD generalized Crewther  relation (GCR), which 
connects  the 
flavor  non-singlet contributions to the Adler and Bjorken polarized 
sum rule   functions, is investigated   at the  $\mathcal{O}(a^4_s)$ 
level of perturbation theory. It is known that 
in the gauge-invariant renormalization $\rm{\overline{MS}}$-scheme  this property holds in the QCD GCR  at least at this order.  To study whether this   factorization property is true  in all gauge-invariant schemes, we consider 
the $\rm{MS}$-like schemes in QCD and the 
QED-limit of the GCR  in the $\rm{\overline{MS}}$-scheme and in  
two other  gauge-independent subtraction schemes, namely in 
 the momentum $\rm{MOM}$ and the on-shell $\rm{OS}$  schemes. 
In these schemes we  confirm the existence of the $\beta$-function 
factorization  
in the QCD and QED variants of the GCR. The problem  of the possible  
$\beta$-factorization     in the  gauge-dependent 
renormalization schemes in QCD
is studied.  To investigate this problem  we consider 
the gauge non-invariant $\rm{mMOM}$ and  $\rm{MOMgggg}$-schemes. We demonstrate that in the $\rm{mMOM}$ scheme at the  $\mathcal{O}(a^3_s)$ level the $\beta$-factorization is valid  for three values of the gauge parameter 
$\xi$ only, namely for $\xi=-3, -1$ and $\xi=0$. In the $\mathcal{O}(a^4_s)$ order of PT it remains valid only for case of the  Landau gauge $\xi=0$.
 The consideration of these two gauge-dependent schemes for the QCD GCR allows us to conclude that the factorization of RG $\beta$-function will always be implemented in any $\rm{MOM}$-like renormalization schemes with linear covariant gauge at $\xi=0$ and $\xi=-3$  at the  $\mathcal{O}(a^3_s)$ approximation.   
It is demonstrated that 
if factorization property for the  $\rm{MS}$-like schemes is true  in all orders of PT, as theoretically indicated in the several works 
on the subject,  then the factorization will also occur in the arbitrary $\rm{MOM}$-like scheme in the  Landau gauge in all orders
of perturbation theory as well.
}


\maketitle
\flushbottom

\section{Introduction}
The triangle Green function, composed from axial-vector-vector (AVV) fermion 
currents, is one of the most interesting quantities in the modern theory of 
electromagnetic and strong interactions. From experimental point of view 
these studies in different kinematic regimes allow to obtain the important
information on say $\pi^0\rightarrow \gamma\gamma$ decay amplitude and 
measurable formfactors of light mesons. From theoretical point of view 
the detailed investigation 
of the  fundamental consequences, obtained from consideration of the
AVV Green function,  leads to the discovery of axial 
anomaly  
and  property  of its non-renormalizability. As was shown in Ref.\cite{Schreier:1971um}  
the result of application of the conformal symmetry (CS)  transformations 
to the triangle diagram with one flavor non-singlet (NS) axial
 and two vector fermion currents  without the internal gauge particles coincides with result, obtained at the lowest order of perturbation theory for triangle diagram, which 
is proportional to $\pi^0\rightarrow\gamma\gamma$ decay amplitude (and the number of the quark colors $N_c$ in particular). In the work  \cite{Crewther:1972kn} within the massless 
quark-parton model it was  proved that in the Born 
approximation the application of the 
operator product expansion (OPE) approach to the AVV triangle diagram in this 
CS limit leads  to the following identity:
\begin{equation}   
\label{Crewther}
D^{NS}_{Born}C_{Bjp, Born}^{NS} = 1~.
\end{equation}
Here $D^{NS}_{Born}$ is the Born approximation of the flavor NS contribution  to 
the normalized   Euclidean characteristic of the 
$e^+e^-\rightarrow \gamma\rightarrow  {hadrons}$ process, namely 
the Adler function $D(a_s)$, which has the following general 
massless renormalized  PT expression:
\begin{equation}
\label{DEM}
D(a_s)=N_c\bigg(\sum\limits_{f}Q^{2}_f\; D^{NS}(a_s)+\bigg(\sum\limits_{f}Q_{f}\bigg)^2D^{SI}(a_s)\bigg)~,
\end{equation}
 where the number of colors  $N_c$ is identical to  the dimension of the 
quark representation of the Lie algebra of the
$SU(N_c)$ color group, $Q_f$ is the electric charge of the  active quark with flavor $f$,  $a_s=\alpha_s/\pi$ and $\alpha_s$ is the QCD coupling constant, $D^{SI}(a_s)$ is the singlet contribution 
to the Adler function, that begins to manifest itself 
from the $\mathcal{O}(a_s^3)$ level \cite{Gorishnii:1990vf}. 

The whole Adler function  
is related 
to the   measurable in the Minkowski  region
characteristic of the electron-positron annihilation to hadrons
process 
$R$-ratio:
\begin{equation} 
\label{R}
R(s)=\frac{\sigma(e^+e^-\rightarrow \gamma\rightarrow  {hadrons})}{\
\sigma_{Born}(e^+e^-\rightarrow \gamma\rightarrow {\mu^+\mu^-})}~, 
\end{equation} 
where 
$\sigma_{Born}(e^+e^-\rightarrow {\mu^+\mu^-})=4\pi\alpha^2_{EM}/(3s)$ 
is the Born massless normalization factor 
with   fixed  expression  of the  QED  coupling constant 
$\alpha_{EM}\approx 1/137$.  
The Adler function and the $R$-ratio 
are related 
through  the following
K\"{a}llen--Lehmann dispersion representation
\begin{equation}
\label{integral}
D(Q^2)=Q^2\int\limits_0^{\infty}ds \frac{R(s)}{(s+Q^2)^2}~,
\end{equation}
and begin to differ in the Minkowski region from  three-loop level due to the effects of the analytical continuation, studied in
 Refs.\cite{Kataev:1995vh, Bakulev:2010gm,  Nesterenko:2017wpb}:
\begin{equation}
R(s)=D(s)-\frac{\pi^2}{3}d_1\beta^2_0a^3_s(s)-\pi^2\bigg(d_2\beta^2_0+\frac{5}{6}d_1\beta_1\beta_0\bigg)a^4_s(s)+\mathcal{O}(a^5_s). 
\end{equation}
The second term $C^{NS}_{Bjp, Born}$ in the l.h.s. of Eq.(\ref{Crewther}) is the Born approximation for the  NS  contribution to the normalized  
characteristic of 
the pure Euclidean process of deep inelastic scattering (DIS) of polarized 
 leptons  on nucleons, namely to the  Bjorken polarized  sum rule, 
which is 
defined as 
\begin{equation} 
\label{Bjp}
\int\limits_0^1 \bigg(g^{lp}_1(x, Q^2)-g^{ln}_1(x, Q^2)\bigg)dx =
\frac{1}{6}\bigg\vert\frac{g_A}{g_V}\bigg\vert
C_{Bjp}(Q^2)~.
\end{equation}
Here $g_1^{lp}(x,Q^2)$ and $g_1^{ln}(x,Q^2)$ are the structure functions of 
these  deep-inelastic scattering  
processes,  which characterize   the spin distribution of quarks 
and gluons inside nucleons, $g_A$ and $g_V$ are the  
 axial and vector neutron $\beta$-decay constants with 
 $g_A/g_V = -1.2723\pm 0.0023$ correspondingly \cite{Patrignani:2016xqp}. The application 
of CVC hypothesis leads to the result   
$g_V$=1. Its application is
sometimes  silently assumed in the definition of the r.h.s. of 
Eq.(\ref{Bjp}). 
The   coefficient function of the  polarized  Bjorken sum rule 
$C_{Bjp}$ can be represented in the following form:
\begin{equation}
C_{Bjp}(a_s)=C^{NS}_{Bjp}(a_s)+N_c\sum_f Q_f C^{SI}_{Bjp}(a_s)~,
\end{equation}
where $C^{SI}_{Bjp}$ is the singlet contribution,   which appears first at 
the  $\mathcal{O}(a^4_s)$ level of PT \cite{Larin:2013yba} with the coefficient, analytically evaluated in 
Ref.\cite{Baikov:2015tea}. Note, that since in this 
 work our main aim is to study theoretical relations between 
coefficients of the PT series for  the $D^{NS}(a_s)$ and $C^{NS}_{Bjp}(a_s)$ 
functions, 
which follows from the  CS limit and its violation by the procedure 
of renormalizations, we neglect in  
Eq.(\ref{DEM}) and (\ref{Bjp}) the  
massive and massless $\mathcal{O}(1/Q^{2k})$ contributions with  
$k\geq 1$. Indeed, these effects    
are the extra  sources of the violation of the CS in QCD.

At the second  stage  of theoretical  studies it is interesting  to 
learn  what will 
happen with the Crewther relation (\ref{Crewther}) in QED-limit 
in the case when the  AVV triangle amplitude 
will contain internal propagators of photons without internal fermion loops.  
In the work of \cite{Adler:1973kz} it was correctly assumed that these insertions   
do not renormalize the AVV triangle graph.
At two-loop level this assumption was confirmed  in Ref.\cite{Jegerlehner:2005fs} by the demonstration of the cancellations of the $\mathcal{O}(\alpha_s)$ corrections
 to the AVV triangle Green function in the most general kinematics.
 In the case of perturbative quenched QED (pqQED)  
 the application of the 
theoretically motivated surmise of Ref.\cite{Adler:1973kz} 
leads to the conclusion that     
the Crewther relation (\ref{Crewther}) should be rewritten 
at the two-loop level as  
\begin{equation}   
\label{pqQED}
D^{NS}(a)C_{Bjp}^{NS}(a) = 1~,
\end{equation}
where $a=\alpha/\pi$ is a  fixed scale-independent QED   
coupling constant. 
Since at this time the two-loop and  three-loop pqQED expressions   
for the 
photon vacuum polarization function  were  
analytically 
calculated in  Refs.\cite{Jost-Luttinger}, \cite{Rosner:1967zz},  
 the application of 
Eq.(\ref{pqQED}) allowed the  authors of  Ref.\cite{Adler:1973kz}
to   predict the analytical expression for the $\mathcal{O}(a^2)$ corrections 
to the  $C_{Bjp}^{NS}(a)$-function  in pqQED.
 It was observed  in 
Ref.\cite{Kataev:2008sk} that these results are in agreement with  pqQED limit 
of the $SU(N_c)$  leading-order (LO) and next-to-leading order (NLO)  QCD results, 
obtained previously in Refs.\cite{Kodaira:1978sh} and \cite{Gorishnii:1985xm} correspondingly.
   
In all orders of pqQED the validity of the  relation (\ref{pqQED})  was 
proved in Ref.\cite{Kataev:2013vua}, where it was clarified that a fixed 
scale-independent coupling constant $a$  should be 
considered as the unrenormalized (bare) QED coupling in the CS limit, which is  realized when the   
charge renormalization constant of QED  
is fixed to unity in all orders of perturbation 
theory, namely  $Z_3=1$\footnote{Note, that the consideration of the  consequences of Eq.(\ref{pqQED}), performed in 
Ref.\cite{Kataev:2008sk},   
 turned out to be rather important 
for  the first confirmation of the validity of the complicated analytical  
calculation of the $\mathcal{O}(\alpha_s^4)$ contribution to $D^{NS}(a_s)$ 
function 
in QCD \cite{Baikov:2008jh}, or to be more precise,  
of the term, containing unexpected previously  
transcendental $\zeta_3$-contribution to  its QED limit, 
firstly discovered  in Ref.\cite{Baikov:2008cp}.}. 
 In   the work \cite{Kataev:2013vua}
it was also explained how  to define this pqQED  limit, which is equivalent 
to the language of the investigated some time ago 
phenomenologically 
unrealized but theoretically important   
finite QED program \cite{Johnson:1967pk}, 
used  in the process of the analytical four-loop computations 
of Ref.\cite{Broadhurst:1999he}, which confirmed the analytical expression of the  
 scheme-independent four-loop contribution to the 
renormalization group QED $\beta$-function 
in the $\rm{MS}$-like and $\rm{MOM}$-schemes, previously evaluated 
in Ref.\cite{Gorishnii:1990kd}
by direct diagram-by-diagram calculations.

However, if one will now consider massless limit 
of  the based on $SU(N_c)$ color group renormalizable QCD or  
of  the realized in nature 
real  QED (which presumes the account of the coupling constant renormalization 
by the ultraviolet divergent expression of the photon renormalization constant 
 $Z_3$),  the Crewther relation, written down in the form of either  (\ref{Crewther}) 
or (\ref{pqQED}), should be modified. The theoretical  
reason for  this modification is related to  the appearance inside  
 AVV triangle graph   of the
internal divergent  subgraphs, which should be 
renormalized. 
  It is known that  the procedure of renormalizations violates the 
CS  and  leads to the appearance of the 
conformal anomaly  \cite{Crewther:1972kn}, \cite{Chanowitz:1972vd},
 which as demonstrated  almost at the same time in  Refs.\cite{Minkowski:1976en,
Adler:1976zt, Collins:1976yq, Nielsen:1977sy} is fixed by the perturbative 
expression of the RG $\beta$-function, coinciding with anomalous dimension of the trace of energy momentum tensor.  In the case of $SU(N_c)$  
QCD it  is usually  evaluated 
in the related to dimensional 
 regularization \cite{tHooft:1972tcz, Bollini:1972ui}  
variant  of the  MS-like schemes  \cite{tHooft:1973mfk}, namely  in the 
$\rm{\overline{MS}}$-scheme \cite{Bardeen:1978yd} \footnote{Its more 
formal  definition is given in  Ref.\cite{Kataev:1988sq}.}.   
This   RG $\beta$-function is responsible for the evolution of the corresponding coupling constant  $a_s(\mu^2)=\alpha_s(\mu^2)/\pi$
and can be  defined as:
\begin{equation}  
\label{RGbeta}
\beta(a_s)=\mu^2\frac{da_s}{d\mu^2}=-\sum_{i\geq 0}\beta_{i}a_s^{i+2}~.
\end{equation}
 
During rather long period  it was unclear how 
the violation of the CS will manifest itself in high orders of the PT  
for  the QCD generalization of the Crewther (GCR) relation in the 
$\rm{\overline{MS}}$-scheme. 
 As   was discovered in Ref.\cite{Broadhurst:1993ru} in this gauge-independent 
scheme  
 the Crewther relation receives additional conformal symmetry breaking 
contribution $\Delta_{csb}(a_s)$, first appearing at the 
 $\mathcal{O}(a_s^2)$ level:
\begin{equation}
\label{GCrew}
D^{NS}(a_s)C_{Bjp}^{NS}(a_s)=
1+\Delta_{csb}(a_s)~, 
\end{equation} 
Moreover, in Ref.\cite{Broadhurst:1993ru} it was shown that 
at least at the $\mathcal{O}(a_s^3)$ level  this extra term can be written 
down in the following factorized form:  
\begin{equation}
\label{BK}
\Delta_{csb}=\bigg(\frac{\beta(a_s)}{a_s}\bigg)K(a_s)~,
\end{equation}
here $\beta(a_s)$ is the two-loop expression for the defined in 
Eq.(\ref{RGbeta}) $\beta$-function of $SU(N_c)$-theory, analytically  
calculated in the $\rm{\overline{MS}}$-scheme
at the one-loop level in Refs.\cite{Gross:1973id, Politzer:1973fx} 
and at the two-loop level in Refs.\cite{Jones:1974mm, Caswell:1974gg, 
 Egorian:1978zx}. The  
function $K(a_s)$ can be represented as the polynomial in powers of 
$a_s$: 
\begin{equation} 
\label{function}
K(a_s)=\sum_{i\geq 1} K_{i}a_s^{i}
\end{equation} 
with the depending on the $SU(N_c)$ 
group structures coefficients $K_i$, computed in Ref.\cite{Broadhurst:1993ru} for $i=1,2$. Its  first term $K_1$  depends only on the defined in the fundamental 
representation  quadratic  Casimir operator  $C_F$, while $K_2$ is expressed 
through  $C_F^2$, $C_FC_A$ and $C_FT_Fn_f$ color structures, 
where $C_A$ is the Casimir operator in the adjoint representation of the Lie algebra, $T_F$ is the Dynkin index and $n_f$ is the number of active 
quark flavors. These coefficients  were obtained from Eqs.(\ref{GCrew}-\ref{function})
due to the knowledge of the  
$SU(N_c)$ structure of the  $\mathcal{O}(a_s^3)$ approximations for $D^{NS}(a_s)$ 
and $C_{Bjp}^{NS}(a_s)$ functions in the $\rm{\overline{MS}}$-scheme  and the two-loop analytical 
expression for the  QCD $\beta$-function\footnote{ The appearance of the term proportional to $\beta_0K_1$  
in Eq.(\ref{BK}) is in agreement with the performed  later on in Ref.\cite{Mondejar:2012sz}  
explicit analytical  calculations of the 
three-loop contributions to the AVV correlator, which contain the internal sub-diagrams  
responsible for the running of the coupling constant and  the
emergence of the $\beta_0$-coefficient.}. 

The corresponding  expression for $D^{NS}(a_s)$ was known thanks to the    
analytical calculations of the  
leading-order (LO)   $\mathcal{O}(a_s)$-correction 
\cite{Zee:1973sr, Appelquist:1973uz}, the next-to-leading order (NLO) 
 $\mathcal{O}(a_s^2)$-correction \cite{Chetyrkin:1979bj, Celmaster:1979xr}
(which agrees with the numerical result for this contribution, independently
obtained in Ref.\cite{Dine:1979qh}) and next-to-next-to-leading order (NNLO)  $\mathcal{O}(a_s^3)$-correction, evaluated in     
Ref.\cite{Gorishnii:1990vf} and confirmed in Ref.\cite{Surguladze:1990tg} and 
later on in  Ref.\cite{Chetyrkin:1996ez} using a bit different technique.
The analytical $\mathcal{O}(a_s^3)$ approximation for the 
$C_{Bjp}^{NS}(a_s)$ function in the $\rm{\overline{MS}}$-scheme, which was used 
in Ref.\cite{Broadhurst:1993ru}, included the information 
on the obtained in Ref.\cite{Kodaira:1978sh} 
$\rm{LO}$ corrections and the  $\rm{NLO}$ and $\rm{NNLO}$-terms, 
evaluated in  Ref.\cite{Gorishnii:1985xm}
and  Ref.\cite{Larin:1991tj} respectively.  

Seventeen years later  the validity of the 
 $\beta$-factorization  property in  $\Delta_{csb}$ term of 
Eq.(\ref{GCrew}), discovered in Ref.\cite{Broadhurst:1993ru},   
 was detected at the $\mathcal{O}(a_s^4)$ level 
 in  Ref.\cite{Baikov:2010je} after getting 
 analytical four-loop $SU(N_c)$  expressions for  the  
$D^{NS}(a_s)$-function (confirmed recently by the 
independent calculation of Ref.\cite{Herzog:2017dtz}) 
and for the  $C_{Bjp}^{NS}(a_s)$-function,  and 
taking into account the explicit expression  
for  the three-loop $SU(N_c)$ RG  $\beta$-function, computed in Ref.\cite{Tarasov:1980au} and 
confirmed in Ref.\cite{Larin:1993tp} 
in  the $\rm{\overline{MS}}$-scheme.
The presented  in Ref.\cite{Baikov:2010je} $\mathcal{O}(a^4_s)$ expression for the conformal 
symmetry breaking term contains the third coefficient
$K_3$ of Eq.(\ref{function}), which  is composed of six color structures, namely 
 $C_F^3, C^2_FC_A, C_FC_A^2, C^2_FT_Fn_f, C_FC_AT_Fn_f$ and $C_FT^2_Fn^2_f$. 

Theoretical arguments in favor of the validity of the $\beta$-factorization 
property 
in the  $\overline{\rm{MS}}$-scheme  
were given in  Ref.\cite{Gabadadze:1995ei} in all orders of PT. 
 At more solid theoretical level this statement was analyzed in the work 
 \cite{Crewther:1997ux} and in the review \cite{Braun:2003rp}.     
However, the detailed understanding and proof  
of the origin of the existence of this 
fundamental  feature of the GCR is still absent.

In this work we  study this problem at the four-loop level using the  method of theoretical    
experiment and   
try to shed some light on the origin of  this feature 
analyzing scheme-dependence  of the 
analytical expression of the GCR  in QCD and QED.     

We start our analysis from the  consideration of  the PT theoretical QCD  expressions for the  NS contributions 
$D^{NS}$ and $C_{Bjp}^{NS}$ 
to the Adler and Bjorken polarized sum rule  functions, which are defined as:
\begin{subequations}
\begin{eqnarray}
\label{Dnsexp}
D^{NS}(a_s)&=&1+\sum_{k\geq 1} d_k a_s^k~, \\ 
\label{Cnsexp}
C_{Bjp}^{NS}(a_s)&=&1+\sum_{k\geq 1} c_ka_s^k~. 
\end{eqnarray}
\end{subequations}
It is already  known that the property of the factorization of the 
$SU(N_c)$ RG $\beta$-function in the GCR is scheme-dependent even within the framework of a gauge-invariant renormalization schemes.
Indeed, as was shown in Ref.\cite{Garkusha:2011xb},  the transformation to the  considered  in 
Ref.\cite{tHooft:1977xjm} 't Hooft scheme with its RG $\beta$-function, 
which contains two nonzero scheme-independent PT coefficients only (the rest are assumed to be zero by finite renormalizations of charge), spoils the 
property of pure $\beta$-factorization in the GCR. In view 
of this the natural question arises whether there do exist 
theoretical requirements to the choice of the ultraviolet subtraction schemes, 
which provide the realization of the fundamental property of the $\beta$-factorization in the GCR,  associated with the manifestation of the conformal symmetry breaking.

In Sec.2 we  study this question in two renormalized gauge theories, 
namely in  $SU(N_c)$ QCD and in QED, based on the Abelian $U(1)$ gauge group.
In this section we  arrive to the conclusion that the  {\it sufficient} 
condition for factorization of the RG $\beta$-function in the GCR at the 
fourth order of PT at least is the choice of the {\it gauge-invariant} 
renormalization scheme for evaluating PT  expressions for the $D^{NS}$, 
$C_{Bjp}^{NS}$-functions and of  the corresponding RG $\beta$-function as well.
We clarify the features of the considered analytical PT expressions 
for the $D^{NS}$ and $C_{Bjp}^{NS}$-functions in various gauge-independent schemes, including the consideration of 
the $\rm{\overline{MS}}$-results,  
discussed previously in Refs.\cite{Broadhurst:1993ru},
\cite{Kataev:2010du, Brodsky:2011ta, Kataev:2014jba, Cvetic:2016rot, Kataev:2016aib, Mikhailov:2016feh} 
and in Refs.\cite{Brodsky:1994eh, Brodsky:1995tb},
\cite{Baikov:2010je}, based on the effective 
charges approach, developed in Refs.\cite{Krasnikov:1981rp, Grunberg:1982fw}.
Remind that the class of gauge-invariant schemes includes not only the original  $\rm{MS}$-scheme 
 and 
widely used in QCD  $\rm{\overline{MS}}$-scheme, but  also
formulated for simplifying multi-loop  calculations G-scheme 
\cite{Chetyrkin:1980pr} and applied at high-order  
computations 
in the chiral perturbation theory   $(\rm{\overline{MS}+1})$-scheme 
\cite{Amoros:1999dp}. 
All these  schemes  
are the elements of the considered in Ref.\cite{Mojaza:2012mf}
infinite set of $\delta$-Renormalization gauge-independent MS-like  schemes.

In the 
QED studies to be presented in this work 
we  use  three gauge-invariant schemes only, namely the  $\rm{\overline{MS}}$-scheme, 
the momentum subtraction (MOM) scheme and the physical-motivated on-shell (OS) scheme.  
 Also  we consider
the scheme dependence of the coefficients $d_k$ and $c_k$, determined in Eqs.(\ref{Dnsexp}) and  (\ref{Cnsexp}),  within  MS-like schemes up to four-loop level. 
After this we examine
the  scheme dependence of the coefficients $K_i$ of the conformal symmetry breaking term in Eq.(\ref{function}) in 
this class of the gauge-invariant schemes.

In Sec.3 we first investigate the question of the possibility of factorization of the RG $\beta$-function in the generalized Crewther relation in 
the \textit{gauge non-invariant} renormalization schemes. For this purpose
we turn to consideration of the popular at present the gauge-dependent subtraction scheme, called the miniMOM ($\rm{mMOM}$) scheme, introduced in Ref.\cite{vonSmekal:2009ae}. We calculate the $\mathcal{O}(a^4_s)$ expression for the NS flavor contribution to the Adler and Bjorken polarized sum rule functions and we find the improved convergence of these  PT series for the  cases of $n_f=3,4,5$ in comparison with the  similar results, obtained in the 
$\rm{\overline{MS}}$-scheme, especially for the Bjorken coefficient function.

In Sec.4 we present an exact recipe 
 for determination of those values of the gauge parameter, for which the  $\beta$-factorization is performed in the GCR at the  $\mathcal{O}(a^3_s)$ and $\mathcal{O}(a^4_s)$  levels. By explicit calculation we demonstrate that in the $\rm{mMOM}$-scheme the gauges  $\xi=0, \; -1,\; -3$ are distinguished from an infinite set of values of gauge parameter $\xi$ at the   $\mathcal{O}(a^3_s)$ level of PT, because exactly for these values the factorization of $\beta$-function in the GCR is fulfilled. At  
the $\mathcal{O}(a^4_s)$ level the factorization property holds in the $\rm{mMOM}$-scheme in the  Landau gauge only. If we fix $\xi=-1$ or $\xi=-3$, 
we observe at this  level  the partial factorization only (see Appendix B). Then  we consider another gauge-dependent renormalization scheme, namely 
the $\rm{MOMgggg}$-scheme, and 
discover definite similarities  between these  
completely different MOM-type  schemes. We 
find that  at  $\xi=0$ and $\xi=-3$ the  $\beta$-factorization property holds in the $\mathcal{O}(a^3_s)$ approximation in the 
$\rm{MOMgggg}$-scheme as well.  
 Based on these at the  first  glance  surprising  results (indeed, it is not immediately clear why the  values $\xi=0$ and $\xi=-3$ are  highlighted), we conclude that in the $\mathcal{O}(a^3_s)$ approximation the factorization of the  $\beta$-function is always possible in an  arbitrary 
$\rm{MOM}$-like renormalization scheme with linear covariant gauges  $\xi=0$ and $\xi=-3$.  Moreover, we prove that if the $\beta$-factorization is valid in $\rm{MS}$-like schemes in all orders of PT, then in the  $\rm{MOM}$-like schemes with the  Landau gauge this factorization property persists in all orders as well. 

\section{Scheme-dependence of the GCR in the gauge-invariant schemes}
\subsection{Scheme-dependence of the PT series in the  $SU(N_c)$ QCD}
Consider first the case of $SU(N_c)$ theory. In this extended version of QCD the most effective method 
of performing high-order PT  calculations is based on the application of dimensional regularization the class of the MS-like gauge-invariant  
ultraviolet  renormalization schemes.  Let us find out how the coefficients of the physical 
quantities discussed above in Eqs.(\ref{Dnsexp}) and (\ref{Cnsexp}) 
change under the transition from one  to another scale 
$\mu\rightarrow\tilde{\mu}$ in this class of  schemes.
It is known that the transformation from the $\rm{\overline{MS}}$-scheme to another 
representative of the class of $\rm{MS}$-like schemes (or following terminology of 
 Ref.\cite{Mojaza:2012mf}    $R_{\delta}$-schemes) can be accomplished by the 
proper  change of the renormalization scale. The transition from the  
$\rm{\overline{MS}}$ to $\rm{MS}$ requires  
following shift  $\tilde{\mu}^2=\mu^2\; {\rm{exp}}(\log 4\pi -\gamma_E)$, 
where $\gamma_E\approx 0.577$ is the constant of Euler--Mascheroni.    
This immediately implies that the PT coefficients  for the  Adler and Bjorken polarized sum rule functions in the  $\rm{MS}$-scheme will contain 
additional terms, proportional to the absent in the $\rm{\overline{MS}}$-scheme transcendental   
($\log 4\pi$-$\gamma_E$)-factor. Indeed, the solution  of the RG equation (\ref{RGbeta}) at 
the  $\mathcal{O}(a^3_s)$-level       
\begin{equation}
\hspace{-0.3cm}
\label{tilde}
a_s(\tilde{\mu}^2)=a_s(\mu^2)\bigg(1-\beta_0{\rm{L}}a_s(\mu^2)+(\beta^2_0{\rm{L}}^2-\beta_1{\rm{L}})a^2_s(\mu^2)+(-\beta^3_0{\rm{L}}^3+\frac{5}{2}\beta_0\beta_1{\rm{L}}^2-\beta_2{\rm{L}})a^3_s(\mu^2)\bigg)
\end{equation}
depends on  the   RG logarithm term   ${\rm{L}}=\log \tilde{\mu}^2/\mu^2$.  
 Using the property of the  RG-invariance of the $D^{NS}(a_s)$-function and 
taking into account Eq.(\ref{tilde}),   
we obtain the following relations reflecting the transformation law of the  coefficients $d_k$ at $\mu\rightarrow\tilde{\mu}$ in the  $\rm{MS}$-like schemes:
\begin{subequations}
\begin{eqnarray}
\label{d1}
\tilde{d}_1&=&d_1~, \\
\label{d2}
\tilde{d}_2&=&d_2+\beta_0 d_1{\rm{L}}~, \\
\label{d3}
\tilde{d}_3&=&d_3+(\beta_1d_1+2\beta_0d_2){\rm{L}}+\beta^2_0d_1{\rm{L}}^2~, \\
\label{d4}
\tilde{d}_4&=&d_4+(\beta_2d_1+2\beta_1d_2+3\beta_0d_3){\rm{L}}+\bigg(\frac{5}{2}\beta_0\beta_1d_1+3\beta^2_0d_2\bigg){\rm{L}}^2+\beta^3_0d_1{\rm{L}}^3~.
\end{eqnarray}
\end{subequations}
It is obvious that coefficients $c_k$ in Eq.(\ref{Cnsexp}) for 
$C_{Bjp}^{NS}(a_s)$  function will obey  the same transformation laws after the  replacement of $d_k$ by $c_k$.

Let us now study the  scheme-dependence of the 
GCR, defined by  Eqs.(\ref{GCrew}-\ref{BK})
in the case of  application of the gauge-invariant $\rm{MS}$-like schemes.
Substituting the $\mathcal{O}(a_s^4)$ PT approximations  for $D^{NS}(a_s)$
and $C^{NS}(a_s)$-functions, given in Eqs.(\ref{Dnsexp}-\ref{Cnsexp}), into the GCR, presented in Eqs.(\ref{GCrew}-\ref{BK}), we get the following 
relations:
 \begin{subequations}
\begin{eqnarray}
\label{d1c1} 
d_1+c_1&=&0~,\\
\label{bK1}
d_2+c_2+d_1c_1&=&-\beta_0K_1~, \\
\label{bK2} 
d_3+c_3+d_1c_2+c_1d_2&=& -\beta_1K_1-\beta_0K_2~, \\
\label{bK3}
d_4+c_4+d_1c_3+c_1d_3+d_2c_2&=&-\beta_2K_1-\beta_1K_2-\beta_0K_3~,
\end{eqnarray}
\end{subequations}
where $\beta_i$ are the coefficients of the  $SU(N_c)$ RG $\beta$-function, 
defined in Eq.(\ref{RGbeta}).  
 The nullification of  Eq.(\ref{d1c1}) is the consequence of the 
conformal symmetry and it reflects the feature of the absence of the   $\mathcal{O}(a_s)$-corrections to 
 the    CSB term   
$\Delta_{csb}$.  The expressions (\ref{bK1}-\ref{bK3}) are valid when the  
property of  factorization of the  $\beta$ function in the GCR 
is true\footnote{Note that due to the absence of 
overall minus in the definition of the QED 
RG $\beta$-function in Eq.(\ref{RGbeta}) the   
QED analogues of these formulas contain pluses
instead  minuses in r.h.s.  of these expressions.}. 

Now everything is ready to study the scheme-dependence of the coefficients $K_i$, included in the CSB term, within the class of $\rm{MS}$-like schemes. Using equations (\ref{d1}-\ref{d4}),  the relations (\ref{d1c1}-\ref{bK3}) and taking into account that these formulas remain valid for any scale\footnote{Remind that transformations $\mu\rightarrow\tilde{\mu}$ do not affect the expressions for the coefficients of the RG $\beta$-function of Eq.(\ref{RGbeta}) in $\rm{MS}$-like schemes.},   
we obtain:  
\begin{subequations}
\begin{gather}
\label{tK1}
\tilde{K}_1=K_1~, \\
\label{tK2}
\tilde{K}_2=K_2+2\beta_0K_1{\rm{L}}~,\\
\label{tK3}
\tilde{K}_3=K_3+3(\beta_1K_1+\beta_0K_2){\rm{L}}+3\beta^2_0K_1{\rm{L}}^2~. 
\end{gather}
\end{subequations}
Note, that  the coefficient $K_1$ is invariant in the  $\rm{MS}$-like schemes and the remaining coefficients $K_2$, $K_3$ are not  
$\rm{MS}$-like schemes invariants. The expressions (\ref{tK1}-\ref{tK3}) coincide with the results,  obtained recently in Ref.\cite{Shen:2016dnq}.
It should be emphasized  that the scale transformations  {\it do not violate}
the property of the factorization of the $SU(N_c)$ $\beta$-function in the PT expression for the CSB term $\Delta_{csb}(a_s)$ of 
the GCR (\ref{GCrew}) within  the gauge-independent $\rm{MS}$-like schemes, but only modify expressions for coefficients $K_i$ according to formulas (\ref{tK1}-\ref{tK3}). This gives us the idea that the gauge-invariance of the 
subtraction schemes  is the {\it sufficient}  condition for the $\beta$-factorization in the CSB massless perturbative term 
of the GCR. To study the status of this 
statement we  extend the considerations, reported in  Ref.\cite{GK2012},  and    
analyze  the GCR in QED, where the number of widely applicable  gauge-independent schemes is larger than in QCD.  In the process of these 
considerations  we use three concrete renormalization schemes, namely for $\rm{\overline{MS}}$, momentum subtraction ($\rm{MOM}$) and on-shell ($\rm{OS}$) gauge-invariant   schemes.

Let us now to move aside slightly  and consider the 
$\{\beta\}$-expansion approach \cite{Mikhailov:2004iq} for representing PT  coefficients of  
the RG invariant quantity, namely for $d_k$-coefficients of the  PT 
 $D^{NS}(a_s)$-series, proposed  and used in Ref.\cite{Mikhailov:2004iq}, which is the high-order PT QCD generalization of the suggested $\rm{\overline{MS}}$-version of the BLM scale-fixing approach in   Ref.\cite{Brodsky:1982gc}\footnote{For the discussion on the scheme-dependence 
of the original BLM procedure within QCD see e.g. \cite{Chyla:1995ej}.}.
 
At $Q^2=\mu^2$ the $\{\beta\}$-expanded representations for $d_k$ have  the 
following form \cite{Mikhailov:2004iq}: 
\begin{subequations}
\hspace{-1cm}
\begin{align}
\label{d2a}
d_1&=d_1[0]~,~~~ d_2= d_2[0]+ \beta_0 d_2[1]~, ~~~
d_3=d_3[0]+\beta_0 d_3[1]+\beta_0^2d_3[2] + \beta_1 d_3[0,1]~,  \\ 
\label{d4a}
d_4&=d_4[0] + 
\beta_0\,d_4[1]+ \beta_0^2 d_4[2] + \beta_0^3 d_4[3]+ \beta_1  d_4[0,1] + \beta_1 \beta_0 d_4[1,1] + 
\beta_2 d_4[0,0,1]~. 
\end{align}     
\end{subequations}
In particular, the  $\{\beta\}$-expansion formalism  was   
used for the construction of the QCD multiloop generalization of the 
$\rm{\overline{MS}}$-version of the BLM approach, called the
Principle of Maximal Conformality (PMC)  \cite{Brodsky:2013vpa}.
The idea of applying the term PMC belongs to
 authors of Ref.\cite{Brodsky:2013vpa}, who correctly realized that 
within the class of $\rm{MS}$-like (or $R_{\delta}$)  schemes the $\beta_k$-independent 
contributions  in $\{\beta\}$-expanded coefficients $d_k$, namely 
$d_k[0]$ in Eqs.(\ref{d2a}-\ref{d4a}), are {\it scheme-independent}
and obey the same properties as the coefficients of  Green functions within
studied previously  
finite QED program \cite{Johnson:1967pk}.
As was explained in Ref.\cite{Kataev:2013vua} this property is related 
to the possibility of defining  the CS limit in this Abelian model.

The presented above RG-based expressions (\ref{d1}-\ref{d4}) clarify  the 
scheme-independent 
property of the terms $d_k[0]$ within the class of MS-like schemes. 
Actually, taking into account the $\{\beta\}$-expanded formalism (\ref{d2a}-\ref{d4a})  and expanding relations (\ref{d1}-\ref{d4}) at an appropriate scale in accordance with this formalism, we obtain that 
 the $\beta_k$-dependent terms of Eqs.(\ref{d2a}-\ref{d4a}) depend 
on the choice of the concrete MS-like prescription of fixing 
$\mu$:
 \begin{subequations}
 \begin{eqnarray*}
\tilde{d}_1[0]&=&d_1[0]~, \\
\tilde{d}_2[0]&=&d_2[0]~, ~~~~~~~~~~~~~~~~~~~~~
\tilde{d}_2[1]=d_2[1]+d_1[0]{\rm{L}}~, \\ 
\tilde{d}_3[0]&=&d_3[0]~, ~~~~~~~~~~~~~~~~~~~~~
\tilde{d}_3[1]=d_3[1]+2d_2[0]{\rm{L}}~, \\ 
\tilde{d}_3[0,1]&=&d_3[0,1]+d_1[0]{\rm{L}}~, ~~~~~~~
\tilde{d}_3[2]=d_3[2]+2d_2[1]{\rm{L}}+d_1[0]{\rm{L}}^2~, \\ 
\tilde{d}_4[0]&=&d_4[0]~, ~~~~~~~~~~~~~~~~~~~~~
\tilde{d}_4[1]=d_4[1]+3d_3[0]{\rm{L}}~, \\ \nonumber
\tilde{d}_4[0,1]&=&d_4[0,1]+2d_2[0]{\rm{L}}~, ~~~~~
\tilde{d}_4[2]=d_4[2]+3d_3[1]{\rm{L}}+3d_2[0]{\rm{L}}^2~,\\ 
\tilde{d}_4[0,0,1]&=&d_4[0,0,1]+d_1[0]{\rm{L}}~,\\ 
\tilde{d}_4[1,1]&=&d_4[1,1]+(2d_2[1]+3d_3[0,1]){\rm{L}}+\frac{5}{2}d_1[0]{\rm{L}}^2~,\\ 
\tilde{d}_4[3]&=&d_4[3]+3d_3[2]{\rm{L}}+3d_2[1]{\rm{L}}^2+d_1[0]{\rm{L}}^3
\end{eqnarray*}
\end{subequations}  
The similar relations holds for the $\{\beta\}$-expanded 
coefficients  of PT series of any RG-invariant physical quantity, 
including $\{\beta\}$-expanded 
coefficients $c_k$ of the NS Bjorken polarized sum rule function $C_{Bjp}^{NS}(a_s)$, 
studied from different points 
of view in the works of \cite{Kataev:2010du},\cite{Kataev:2014jba},\cite{Shen:2016dnq},\cite{Deur:2017cvd}. 

 In accordance with the proposals of Ref.\cite{Mikhailov:2004iq},\cite{Brodsky:2013vpa},\cite{Mojaza:2012mf} all $\beta_k$-dependent contributions in Eqs.(\ref{d2a}-\ref{d4a}) should be absorbed 
into the set of scales of the  initially defined in the $\rm{\overline{MS}}$-scheme QCD coupling constant. In view of the scheme-dependence of these $\beta_k$-dependent terms, the scales of the resulting QCD-expansion parameter become scheme-dependent as well and do not respect the CS approximations. Therefore, the term Principle of Maximal Conformality, used in the number of the related works 
on this topic \cite{Brodsky:2011ta},\cite{Mojaza:2012mf},\cite{Shen:2016dnq},\cite{Brodsky:2013vpa, Deur:2017cvd, Ma:2015dxa}, should be used with care. 

Note that in Ref.\cite{Kataev:2014jba} the  NNLO   PMC-type  
procedure to both $D^{NS}(a_s)$ and  $C_{Bjp}^{NS}(a_s)$-functions was applied  in the QCD-type model, based on the  $SU(N_c)$ color 
gauge theory, supplemented with multiplet of SUSY gluino. The 
analytical results for the $\{\beta\}$-expanded expressions
of $d_2$ and $d_3$ coefficients in this model were first obtained in  Ref.\cite{Mikhailov:2004iq}
and confirmed later on in Ref.\cite{Ma:2015dxa}. In Ref.\cite{Cvetic:2016rot} it was demonstrated how to define the 
expressions for  
 $\{\beta\}$-expanded coefficients of the
Adler and Bjorken functions at the $\mathcal{O}(a_s^4)$ level 
in  $SU(N_c)$ QCD without any gluino. The careful QCD $\mathcal{O}(a_s^4)$   reconsideration of the results of Ref.\cite{Kataev:2014jba}  and Refs.\cite{Mojaza:2012mf},\cite{Shen:2016dnq},\cite{Brodsky:2013vpa, Deur:2017cvd, Ma:2015dxa} with taking into account that 
the QCD  anomalous dimension of the photon vacuum
polarization function has its own well-defined $\beta$-expansion structure (for the detailed clarification of this 
point see  Ref.\cite{Kataev:2016aib}) is on the agenda and will be considered elsewhere.

 We will not  
discuss anymore  these  phenomenologically oriented problems, but will return to one of  the  theoretical aims 
of this work, namely  to the study of the scheme (in)dependence of the CSB PT contribution to the GCR in the factorized by RG $\beta$-function form
in the different schemes, including the ones commonly used in QED. In order to analyze this problem we should 
get the $\mathcal{O}(a^4)$ analytical PT approximations for the NS Adler function and Bjorken polarized sum rule function in QED in the 
gauge-invariant renormalization schemes we are interested in.    

\subsection{The NS Adler function in the  $\rm{\overline{MS}}$, $\rm{MOM}$ and $\rm{OS}$ schemes in QED}

In order to obtain the $\mathcal{O}(a^4)$ expression for the non-singlet QED 
Adler function in  the $\rm{\overline{MS}}$ scheme we use the  results of 
Ref.\cite{Baikov:2010je}.  We consider QED with $N$ types of charged 
leptons ($N=1,\; 2,\; 3$ for $e,\; \mu,\; \tau$-leptons correspondingly). 
Fixing $C_F=1$, $C_A=0$, $T_F=1$, $d^{abcd}_F=1$, $d^{abcd}_A=0$, $d_R=1$, $n_f=N$,
  we obtain the following four-loop analytical approximation for the $D^{NS}(a)$-function in the  $\rm{\overline{MS}}$-scheme, which is related through   the 
K\"{a}llen--Lehmann dispersion representation  with the massless PT expression for the total cross-section of   the process  $e^+e^-\rightarrow\gamma\rightarrow  l^+l^-$
 (here $l$ denotes one of the charged leptons $e, \mu$ or $\tau$-lepton):
\begin{subequations}
\begin{align}
\label{r0ms}
D^{NS}_{{\rm{\overline{MS}}}}(a)&=1+\sum\limits_{k=1}^4 d^{{\rm{\overline{MS}}}}_k(a^{{\rm{\overline{MS}}}})^k~, \\ \nonumber
\end{align}
\begin{align}
\label{r1ms}
d^{{\rm{\overline{MS}}}}_1&= \frac{3}{4}~, ~~~
d^{{\rm{\overline{MS}}}}_2=-\frac{3}{32}+\bigg(-\frac{11}{8}+\zeta_3\bigg)N~, \\
d^{{\rm{\overline{MS}}}}_3&=-\frac{69}{128}+
\bigg(-\frac{29}{64}+\frac{19}{4}\zeta_3-
 5\zeta_5\bigg)N+
\bigg(\frac{151}{54}-\frac{19}{9}\zeta_3\bigg)N^2~, \\
\label{r3ms}
d^{{\rm{\overline{MS}}}}_4&=\frac{4157}{2048}+\frac{3}{8}\zeta_3+
\bigg(\frac{689}{384}+\frac{67}{32}\zeta_3-\frac{115}{4}\zeta_5
+\frac{105}{4}\zeta_7\bigg)N \\ \nonumber  
&+\bigg(\frac{5713}{1728}-\frac{581}{24}\zeta_3+\frac{125}{6}\zeta_5
+3\zeta_3^2\bigg)N^2+\bigg(-\frac{6131}{972}+\frac{203}{54}\zeta_3+\frac{5}{3}\zeta_5\bigg)N^3~.
\end{align}
\end{subequations}
The energy behavior of the  QED coupling constant  $a^{{\rm{\overline{MS}}}}=\alpha^{{\rm{\overline{MS}}}}/\pi$
 in  Eq.(\ref{r0ms}) is   determined by the four-loop expression of the corresponding $\beta$-function, which reads
\begin{eqnarray}
\label{betaMSQED}
\beta^{{\rm{\overline{MS}}}}(a^{\rm{\overline{MS}}})&=&\frac{1}{3}N(a^{\rm{\overline{MS}}})^2+\frac{1}{4} N (a^{\rm{\overline{MS}}})^3+
\bigg(-\frac{1}{32}N - \frac{11}{144} N^2\bigg) (a^{\rm{\overline{MS}}} )^4  \\
\nonumber  
 &+&\bigg( -\frac{23}{128} N + \bigg[\frac{95}{864}-\frac{13}{36} \zeta_3\bigg] N^2-\frac{77}{3888} N^3  \bigg) (a^{\rm{\overline{MS}}})^5~.
\end{eqnarray}
Note, that the first two scheme-independent coefficients, included in Eq.(\ref{betaMSQED}),
were obtained in Ref.\cite{GellMann:1954fq} from the two-loop 
approximation of the photon vacuum polarization function  
$\Pi(Q^2/\mu^2,a)$, evaluated  previously  at the same level  in Ref.\cite{Jost-Luttinger}. 
The $N$-dependence of the three-loop coefficient in the $\rm{\overline{MS}}$-scheme was obtained in Ref.\cite{Gorishnii:1987fy}. At $N=1$ it agrees with the 
results of calculations, independently performed in  Refs.\cite{Chetyrkin:1980pr}
and  \cite{Vladimirov:1979zm}. 
The four-loop 
term was obtained in Ref.\cite{Gorishnii:1990kd}. 

To transform the results of Eqs.(\ref{r1ms}-\ref{r3ms})  to the 
$\rm{MOM}$ and $\rm{OS}$-schemes  we use the following RG-based relations   
\begin{eqnarray}
\label{MOM}
\beta^{{\rm{MOM}}}(a^{{\rm{MOM}}})&=&\beta^{{\rm{\overline{MS}}}}(a^{{\rm{\overline{MS}}}})
\;\partial a^{{\rm{MOM}}}/\partial a^{{\rm{\overline{MS}}}}~, \\ 
\label{OS} 
\beta^{{\rm{OS}}}(a^{{OS}})&=&\beta^{{\rm{\overline{MS}}}}(a^{{\rm{\overline{MS}}}})
\;\partial a^{{\rm{OS}}}/\partial a^{{\rm{\overline{MS}}}}~,
\end{eqnarray}    
and take into account  the properties  of the RG-invariance and scheme-independence 
of the  QED invariant charge     
$a/(1+\Pi(Q^2/\mu^2,a))$ or to be more precise of its expression, related to the 
$\rm{\overline{MS}}$, $\rm{MOM}$ and $\rm{OS}$ schemes.
Note also that in the  $\rm{MOM}$ scheme, determined by subtractions of the UV divergences of the photon 
vacuum polarization function at the non-zero Euclidean momentum, the QED $\beta$-function 
coincides  with  the Gell-Mann--Low function,  which governs the energy 
behavior of the 
QED invariant charge. 

The four-loop expressions for the QED $\beta$-functions in the  
$\rm{MOM}$ and $\rm{OS}$ schemes are 
known and have the following form: 
\begin{align}
\hspace{-0.5cm}
 \label{betaMOMQED}
\beta^{{\rm{MOM}}}(a^{{\rm{MOM}}})&=\frac{1}{3}N(a^{{\rm{MOM}}})^2+\frac{1}{4}N(a^{{\rm{MOM}}})^3+\bigg(-\frac{1}{32}N -
\bigg[\frac{23}{72}-\frac{1}{3}\zeta_3\bigg] N^2\bigg) (a^{{\rm{MOM}}})^4  \\ \nonumber
&+
\bigg( -\frac{23}{128} N + \bigg[\frac{13}{32} + \frac{2}{3} \zeta_3 - \frac{5}{3} \zeta_5\bigg] N^2  +
\bigg[\frac{1}{2} -\frac{1}{3} \zeta_3\bigg] N^3 \bigg) (a^{{\rm{MOM}}})^5~, \\
\label{betaOSQED} 
\beta^{{\rm{OS}}} (a^{{\rm{OS}}}) &= \frac{1}{3} N (a^{{\rm{OS}}})^2 + \frac{1}{4} N (a^{{\rm{OS}}})^3 + \bigg(-
\frac{1}{32}N  - \frac{7}{18} N^2\bigg) (a^{{\rm{OS}}})^4 +\bigg(-\frac{23}{128} N  \\ \nonumber
&+\bigg[\frac{1}{48} - \frac{5}{6} \zeta_2 + \frac{4}{3} \zeta_2 \log 2 - \frac{35}{96} \zeta_3\bigg] N^2 +
\bigg[\frac{901}{1296} - \frac{4}{9} \zeta_2 - \frac{7}{96} \zeta_3\bigg] N^3 \bigg) (a^{{\rm{OS}}})^5~. 
\end{align}
The proportional to $N$ scheme-independent contribution to the 
three-loop corrections of 
Eqs.(\ref{betaMOMQED}) and (\ref{betaOSQED}) was calculated long time ago in  
Ref.\cite{Rosner:1967zz}, while the full  three-loop corrections to  
Eqs.(\ref{betaMOMQED}) and (\ref{betaOSQED}) were 
evaluated in Refs.\cite{Gorishnii:1987fy}\footnote{For $N$=1 the obtained in Ref.\cite{Gorishnii:1987fy} 
three-loop analytical result coincides with the one, previously computed in Ref.\cite{Baker:1969an}.}
and \cite{DeRafael:1974iv} correspondingly. The analytical expressions for the four-loop 
contributions to (\ref{betaMOMQED}) and (\ref{betaOSQED}) were calculated in the works 
of \cite{Gorishnii:1990kd} and \cite{Broadhurst:1992za}.

Fixing $Q^2=\mu^2_{\rm{\overline{MS}}}=\mu^{2}_{\rm{MOM}}=m^2_{OS}$ (that reflects 
theoretical freedom in fixing scale parameters of the $\rm{\overline{MS}}$ 
and $\rm{MOM}$-schemes in QED) and taking into account the scheme-independence of the 
constant term of the  single lepton-loop contribution to the photon vacuum 
polarization function included in the QED invariant charge \cite{Broadhurst:1992za, Gorishnii:1991hw, Baikov:2012rr}, 
 and applying the 
results of Eqs.(\ref{betaMSQED}-\ref{betaOSQED}), one can obtain  the following 
transformation relations:
\begin{eqnarray}
\label{aMOMQED}
a^{{\rm{\overline{MS}}}}&=&a^{{\rm{MOM}}}+\bigg(\frac{35}{48} -\zeta_3\bigg)N(a^{{\rm{MOM}}})^3+\bigg(\bigg[-\frac{4}{9}-\frac{37}{24}\zeta_3+\frac{5}{2}\zeta_5\bigg]N \\ \nonumber
&+&\bigg[-\frac{2021}{2592}+\frac{\zeta_3}{2}\bigg]N^2\bigg)(a^{{\rm{MOM}}})^4~, \\ 
\label{aOSQED}
a^{{\rm{\overline{MS}}}}&=&a^{{\rm{OS}}}+\frac{15}{16}N(a^{{\rm{OS}}})^3+\bigg(\bigg[\frac{77}{576}+\frac{5}{4}\zeta_2-2\zeta_2\log 2+\frac{\zeta_3}{192}\bigg]N \\ \nonumber
&+&
\bigg[-\frac{695}{648}+\frac{2}{3}\zeta_2+\frac{7}{64}\zeta_3\bigg]N^2\bigg)(a^{{\rm{OS}}})^4~.
\end{eqnarray}
Using expressions (\ref{r0ms}-\ref{r3ms}), the obtained expansions 
(\ref{aMOMQED}-\ref{aOSQED})  and taking into consideration that the flavor 
NS Adler function is the renormalization group 
invariant quantity, we get  the values of  the coefficients $d_k$ of 
the QED  PT series for  Eq.(\ref{Dnsexp}) in the $\rm{MOM}$-scheme:
\begin{subequations}
\begin{align}
\hspace{-0.9cm}
D^{NS}_{{\rm{MOM}}}(a^{{\rm{MOM}}})&=1+\sum\limits_{k=1}^4 
d^{{\rm{MOM}}}_k(a^{{\rm{MOM}}})^k~, \\
\label{d1mom}
d^{{\rm{MOM}}}_1&= \frac{3}{4}~, ~~~
d^{{\rm{MOM}}}_2=-\frac{3}{32}+\bigg(-\frac{11}{8}+\zeta_3\bigg)N~, \\
d^{{\rm{MOM}}}_3&=-\frac{69}{128}+\bigg(\frac{3}{32}+4\zeta_3-5\zeta_5\bigg)N+\bigg(\frac{151}{54}-\frac{19}{9}\zeta_3\bigg)N^2~, \\
\label{d4mom}
d^{{\rm{MOM}}}_4&=\frac{4157}{2048}+\frac{3}{8}\zeta_3+\bigg(\frac{339}{256}+\frac{9}{8}\zeta_3-\frac{215}{8}\zeta_5+\frac{105}{4}\zeta_7\bigg)N \\ \nonumber
&+\bigg(\frac{275}{384}-\frac{157}{8}\zeta_3+\frac{125}{6}\zeta_5+\zeta^2_3\bigg)N^2+\bigg(-\frac{6131}{972}+\frac{203}{54}\zeta_3+\frac{5}{3}\zeta_5\bigg)N^3~,
\end{align}
\end{subequations}
and in  the physically motivated $\rm{OS}$-scheme:
\begin{subequations}
\begin{align}
\hspace{-0.2cm}
D^{NS}_{{\rm{OS}}}(a^{{\rm{OS}}})&=1+\sum\limits_{k=1}^4
d^{{\rm{OS}}}_k(a^{{\rm{OS}}})^k~, \\ 
\label{d1os}
d^{{\rm{OS}}}_1&= \frac{3}{4}~, ~~~
d^{{\rm{OS}}}_2=-\frac{3}{32}+\bigg(-\frac{11}{8}+\zeta_3\bigg)N~, \\ \nonumber
\end{align}
\begin{align}
d^{{\rm{OS}}}_3&=-\frac{69}{128}+\bigg(\frac{1}{4}+\frac{19}{4}\zeta_3-5\zeta_5\bigg)N+\bigg(\frac{151}{54}-\frac{19}{9}\zeta_3\bigg)N^2~, \\ 
\label{d4os}
d^{{\rm{OS}}}_4&=\frac{4157}{2048}+\frac{3}{8}\zeta_3+\bigg(\frac{55}{32}+\frac{15}{16}\zeta_2+\frac{537}{256}\zeta_3-\frac{115}{4}\zeta_5+\frac{105}{4}\zeta_7-\frac{3}{2}\zeta_2\log 2\bigg)N \\ \nonumber
&+\bigg(-\frac{11}{144}+\frac{\zeta_2}{2}-\frac{17089}{768}\zeta_3+\frac{125}{6}\zeta_5+3\zeta^2_3\bigg)N^2+\bigg(-\frac{6131}{972}+\frac{203}{54}\zeta_3+\frac{5}{3}\zeta_5\bigg)N^3,
\end{align}
\end{subequations}
where $a^{{\rm{MOM}}}$ and $a^{{\rm{OS}}}$ are the QED running coupling constants 
of the $\rm{MOM}$ and $\rm{OS}$ schemes, which at the studied by us level  
are the solutions of the corresponding RG equation, with the 
$\rm{MOM}$ and $\rm{OS}$ four-loop expressions of the QED $\beta$-functions, 
presented  in Eqs.(\ref{betaMOMQED}-\ref{betaOSQED}). 

It is worth emphasizing  that the analytical 
expressions for the coefficients of the  $\mathcal{O}(a)$ and 
$\mathcal{O}(a^2)$ terms 
for the QED expressions of the 
$D^{NS}$-function in the $\rm{\overline{MS}}$, $\rm{MOM}$ and 
$\rm{OS}$-schemes, are the same. This circumstance is a consequence of the presented in Eqs.(\ref{aMOMQED}-\ref{aOSQED}))
relations between QED  running coupling constants of 
 $\rm{\overline{MS}}$, $\rm{MOM}$ and 
$\rm{OS}$-schemes, which
do not  contain    $\mathcal{O}(a)$-corrections in the case, when we choose $\mu^2_{\rm{\overline{MS}}}=\mu^{2}_{\rm{MOM}}=m^2_{OS}$.

Let us comment the observations, which follow from the 
comparison of the analytical  expressions for the $D^{NS}$ coefficients in Eqs.(\ref{r1ms}-\ref{r3ms}), 
 (\ref{d1mom}-\ref{d4mom}) and (\ref{d1os}-\ref{d4os}) in three gauge-invariant schemes
mentioned above.
\begin{enumerate}
\item The scheme-independence of the proportional to $N^{0}$-contributions
to the  the QED  PT expression for the NS Adler function follows from the 
CS, which is valid in the case of consideration of the pqQED approximation 
of the PT series for the RG invariant quantities (for detailed explanation see 
Ref.\cite{Kataev:2013vua}).
\item As explained above, the $\mathcal{O}(a^2)$ QED PT contributions to $D^{NS}$  
are scheme-independent.  
\item The scheme-independence of the leading on $N$ 
corrections to the coefficients of $d_{k}$ are the consequence of the 
scheme-independence of the leading  renormalon contributions to the 
QED $D^{NS}$-function.
\item Note also the scheme-independence of the proportional to $\zeta_3$,  
$\zeta_5$ and 
$\zeta_7$  
high transcendence contributions to the proportional to $N$ terms in the 
PT coefficients  $d_2$, $d_3$ and $d_4$. 
This interesting feature is not yet understood. 
\end{enumerate} 

\subsection{The NS Bjorken function in the   $\rm{\overline{MS}}$, $\rm{MOM}$ and $\rm{OS}$ schemes in QED}

Consider now the $\mathcal{O}(a^4)$ approximations to the    
QED analytical expressions for the non-singlet coefficient function 
of the Bjorken polarized sum rule in three schemes we are interested in. 
Using the inverse $\mathcal{O}(a^4_s)$ $SU(N_c)$ QCD 
 $\rm{\overline{MS}}$-scheme results given in  Ref.\cite{Baikov:2010je},   one can  find the following  analytical expression
  for the 
QED corrections to the 
Bjorken polarized  sum rule function in the $\rm{\overline{MS}}$-scheme:   
\begin{subequations}
\begin{align}
C^{NS}_{Bjp,\;{\rm{\overline{MS}}}}(a^{{\rm{\overline{MS}}}})&=1+\sum\limits_{k=1}^4 c^{{\rm{\overline{MS}}}}_k(a^{{\rm{\overline{MS}}}})^k~, \\ 
\label{c1ms}
c^{{\rm{\overline{MS}}}}_1&=-\frac{3}{4}~, ~~~
c^{{\rm{\overline{MS}}}}_2=\frac{21}{32}+\frac{N}{2}~, \\  \nonumber
\end{align}
\begin{align}
\label{c3ms}
c^{{\rm{\overline{MS}}}}_3&=-\frac{3}{128}+\bigg(-\frac{133}{576}-\frac{5}{12}\zeta_3\bigg)N-\frac{115}{216}N^2~,\\ 
\label{c4ms}
c^{{\rm{\overline{MS}}}}_4&=-\frac{4823}{2048}-\frac{3}{8}\zeta_3+\bigg(\frac{2711}{2304}+\frac{547}{96}\zeta_3-\frac{205}{24}\zeta_5\bigg)N \\ \nonumber
&+\bigg(-\frac{265}{576}+\frac{29}{24}\zeta_3\bigg)N^2+\frac{605}{972}N^3~.
\end{align}
\end{subequations}
 Applying now the relations (\ref{aMOMQED}-\ref{aOSQED}) between 
the QED  running coupling constants of the  $\rm{\overline{MS}}$, $\rm{MOM}$ and $\rm{OS}$ schemes, we  get the following analytical  $\mathcal{O}(a^4)$ PT QED approximations 
for $C^{NS}_{Bjp}$-function in the $\rm{MOM}$ and $\rm{OS}$-schemes:
\begin{subequations}
\begin{align}
C^{NS}_{Bjp,\;{\rm{MOM}}}(a^{{\rm{MOM}}})&=1+\sum\limits_{k=1}^4 c^{{\rm{MOM}}}_k(a^{{\rm{MOM}}})^k~, \\
\label{c1mom}
c^{{\rm{MOM}}}_1&=-\frac{3}{4}~, ~~~ c^{{\rm{MOM}}}_2=\frac{21}{32}+\frac{N}{2}~, \\
c^{{\rm{MOM}}}_3&=-\frac{3}{128}+\bigg(-\frac{7}{9}+\frac{\zeta_3}{3}\bigg)N-\frac{115}{216}N^2~, \\
\label{c4mom}
c^{{\rm{MOM}}}_4&=-\frac{4823}{2048}-\frac{3}{8}\zeta_3+\bigg(\frac{1421}{576}+\frac{133}{24}\zeta_3-\frac{125}{12}\zeta_5\bigg)N \\ \nonumber
&+\bigg(\frac{2951}{3456}-\frac{\zeta_3}{6}\bigg)N^2+\frac{605}{972}N^3~,
\end{align}
\end{subequations}
\begin{subequations}
\begin{align}
\hspace{-0.3cm}
C^{NS}_{Bjp,\;{\rm{OS}}}(a^{{\rm{OS}}})&=1+\sum\limits_{k=1}^4 c^{{\rm{OS}}}_k(a^{{\rm{OS}}})^k~, \\
\label{c1os}
c^{{\rm{OS}}}_1&=-\frac{3}{4}~, ~~~ c^{{\rm{OS}}}_2=\frac{21}{32}+\frac{N}{2}~, \\
c^{{\rm{OS}}}_3&=-\frac{3}{128}+\bigg(-\frac{269}{288}-\frac{5}{12}\zeta_3\bigg)N-\frac{115}{216}N^2~,\\
\label{c4os}
c^{{\rm{OS}}}_4&=-\frac{4823}{2048}-\frac{3}{8}\zeta_3+\bigg(\frac{5315}{2304}-\frac{15}{16}\zeta_2+\frac{4373}{768}\zeta_3-\frac{205}{24}\zeta_5+\frac{3}{2}\zeta_2\log 2\bigg)N \\ \nonumber
&+\bigg(\frac{2215}{1728}-\frac{\zeta_2}{2}+\frac{865}{768}\zeta_3\bigg)N^2+\frac{605}{972}N^3~,
\end{align}
\end{subequations}
Note, that three  from  four noticed in Sec.2.2  properties of the analytical 
structure of the 
PT series for the $D^{NS}(a)$-function are also valid in the case of  the $C^{NS}_{Bjp}(a)$-function, 
considered in the gauge-invariant 
$\rm{\overline{MS}}$, $\rm{MOM}$ and $\rm{OS}$ schemes. 
The fourth observed feature is violated. Indeed, the   
proportional to $N$   
 contributions  to the  $c_2$, $c_3$ and $c_4$ coefficients of Eq.(\ref{Cnsexp})  
do not contain high-transcendent functions $\zeta_3$, $\zeta_5$ and 
$\zeta_7$ 
\footnote{These high-transcendental  terms are 
contained in the proportional to $C_A$ non-abelian contributions to the corresponding 
coefficients of the PT  $SU(N_c)$ expressions for the    $C^{NS}_{Bjp}(a_s)$-function in  
the $\rm{\overline{MS}}$-scheme (see   Ref.\cite{Baikov:2010je}) , which is not studied  in this section.}.
 
\subsection{The QED  generalized Crewther relation in the $\rm{\overline{MS}}$, $\rm{MOM}$ and $\rm{OS}$ schemes}

Having now at hand the analytical $\mathcal{O}(a^4)$ QED results  of Sec.2.2 for the $D^{NS}(a)$-function 
in the $\rm{\overline{MS}}$, $\rm{MOM}$ and $\rm{OS}$ schemes and 
the presented in Sec.2.3 similar PT expressions for the
$C_{Bjp}^{NS}(a)$-function we are able to verify  whether
the the conformal symmetry breaking 
term  $\Delta_{csb}$ of the GCR obeys the property of the 
$\beta$-factorization in the class of  the
 $\rm{MS}$-like schemes only, 
or it is also fulfilled (at least in QED) in other gauge-invariant schemes, namely   $\rm{MOM}$ and $\rm{OS}$ schemes.
To understand this we assume, that in QED the  transformations from $\rm{\overline{MS}}$-scheme 
to  $\rm{MOM}$ and $\rm{OS}$ schemes will not spoil the structure of the GCR given in 
Eqs.(\ref{GCrew}-\ref{function}), which is valid at 
the $\mathcal{O}(a_s^4)$ level in  the $\rm{MS}$-like schemes for sure and 
leads to the Eqs.(\ref{d1c1}-\ref{bK3}), obtained in Sec.2.1. It is clear that Eq.(\ref{d1c1}), which is one of the CS relations, 
encoded in the studied in Ref.\cite{Adler:1973kz}
pqQED variant of the  original Crewther relation, is valid for the 
scheme-independent coefficients $c_1$ and $d_1$ in $\rm{MOM}$ and $\rm{OS}$-schemes.  

Note, that  the used in this work and   defined in 
Eqs.(\ref{betaMSQED}), (\ref{betaMOMQED}), (\ref{betaOSQED}) PT expressions  
for  the QED $\beta$-functions 
in the $\rm{\overline{MS}}$, $\rm{MOM}$ and $\rm{OS}$-schemes differ from 
introduced in Eq.(\ref{RGbeta}) determination of the QCD RG $\beta$-function 
by the overall sign. Keeping this in mind, using the 
given in Sec.2.2 and 2.3 
analytical QED  expressions for the coefficients $d_k$ and $c_k$
(with $1\leq k\leq 4$) in the  $\rm{\overline{MS}}$,  $\rm{MOM}$ and 
$\rm{OS}$-schemes and substituting the corresponding expressions for 
two scheme-independent coefficients  $\beta_0$ and $\beta_1$ of the 
RG QED $\beta$-function and scheme-dependent coefficient $\beta_2$ in 
the  system of equations (\ref{bK1}-\ref{bK3}), 
we obtain the following analytical  expressions for the $K_i$-terms:
\begin{align}
\hspace{-0.4cm}
\label{K1QED}
K^{{\rm{\overline{MS}}}}_1&=K^{{\rm{MOM}}}_1=K^{{\rm{OS}}}_1=-\frac{21}{8}+3\zeta_3~, \\ 
\label{K2QED}
K^{{\rm{\overline{MS}}}}_2&=K^{{\rm{MOM}}}_2=K^{{\rm{OS}}}_2=\frac{397}{96}+\frac{17}{2}\zeta_3-15\zeta_5+\bigg(\frac{163}{24}-\frac{19}{3}\zeta_3\bigg)N~, \\
K^{{\rm{\overline{MS}}}}_3&=\frac{2471}{768}+\frac{61}{8}\zeta_3-\frac{715}{8}\zeta_5+\frac{315}{4}\zeta_7+\bigg(-\frac{7729}{1152}-\frac{917}{16}\zeta_3+\frac{125}{2}\zeta_5+9\zeta^2_3\bigg)N \\ \nonumber
&+ \bigg(-\frac{307}{18}+\frac{203}{18}\zeta_3+5\zeta_5\bigg)N^2~, \\
K^{{\rm{MOM}}}_3&=\frac{2471}{768}+\frac{61}{8}\zeta_3-\frac{715}{8}\zeta_5+\frac{315}{4}\zeta_7+\bigg(-\frac{1793}{144}-\frac{343}{8}\zeta_3+\frac{125}{2}\zeta_5\bigg)N \\ \nonumber
&+ \bigg(-\frac{307}{18}+\frac{203}{18}\zeta_3+5\zeta_5\bigg)N^2~, \\
\label{K3QED}
K^{{\rm{OS}}}_3&=\frac{2471}{768}+\frac{61}{8}\zeta_3-\frac{715}{8}\zeta_5+\frac{315}{4}\zeta_7+\bigg(-\frac{8117}{576}-\frac{391}{8}\zeta_3+\frac{125}{2}\zeta_5+9\zeta^2_3\bigg)N \\ \nonumber
&+\bigg(-\frac{307}{18}+\frac{203}{18}\zeta_3+5\zeta_5\bigg)N^2~.
\end{align} 
These results are convincing us that the factorization property of the 
$\beta$-function in the conformal symmetry breaking term $\Delta_{csb}$ to the 
GCR in QED is valid  at least at the fourth order of massless PT at least in three   
widely used in QED \textit{gauge-invariant} schemes, namely in the  $\rm{\overline{MS}}$ (and therefore
$\rm{MS}$-like),  $\rm{MOM}$ and $\rm{OS}$ schemes. Moreover, we discover that in QED  the first two 
coefficients $K_1$ and $K_2$ in expansion of $\Delta_{csb}$  term are scheme-independent and 
find that the  $K_3$-contributions, computed  
in the  $\rm{\overline{MS}}$, $\rm{MOM}$ and $\rm{OS}$-schemes,  differ from each other in the  
linear in the number of leptons $N$ term 
 only:
\begin{equation}
K^{{\rm{\overline{MS}}}}_3=K^{{\rm{MOM}}}_3+\bigg(\frac{735}{128}-\frac{231}{16}\zeta_3+9\zeta^2_3\bigg)N=K^{{\rm{OS}}}_3+\bigg(\frac{945}{128}-\frac{135}{16}\zeta_3\bigg)N~.
\end{equation} 
Completing this Section we arrive to the 
conclusion that really the factorization of the $\beta$-function in the GCR  
takes place in the \textit{gauge-invariant} renormalization schemes, 
such as $\rm{MS}$-like, $\rm{MOM}$, $\rm{OS}$-schemes in QED and $\rm{MS}$-like schemes in QCD.  However, another important problem arises, namely whether  
\textit{gauge invariance} of the renormalization subtraction schemes is 
a \textit{necessary}  
condition for factorization of the $\beta$-function in the GCR. 
This problem is studied below for the case 
of  QCD with $SU(N_c)$ color gauge group.

\section{May the $\beta$-factorization property manifest itself  
in the generalized Crewther relation in the gauge-dependent schemes in QCD?}

It is known that in QCD  the $\rm{MS}$-like subtractions   schemes
are  distinguished by their  gauge-independence. However, in theoretical 
and phenomenological QCD applications the gauge-dependent $\rm{MOM}$-like schemes 
are used as well \cite{Celmaster:1979km, Braaten:1981dv, Hagiwara:1982ct, Tarasov:1990ps, Jegerlehner:1998zg, Chetyrkin:2000fd}.
Among the number of  $\rm{MOM}$ subtractions schemes one of the most applicable at present
is the  miniMOM ($\rm{mMOM}$) scheme, which was originally  formulated  
in Ref.\cite{vonSmekal:2009ae}. It is widely used in different 
QCD-oriented studies \cite{Gracey:2013sca, Gracey:2014pba, Ryttov:2014nda, Kataev:2015yha, Zeng:2015gha, Ayala:2016zrz, Ayala:2017tco, Ruijl:2017eht},\cite{Herzog:2017dtz}. In this Section we will use this $\rm{mMOM}$-scheme 
in the  $SU(N_c)$ theory to find out 
the constraints imposed by  
 its gauge-dependence on the conditions of existence 
of the fundamental property  
of the $\beta$-function factorization in the GCR.

\subsection{The reminding notes on the  miniMOM scheme} 

The basic requirement, which is lying beyond the definition of 
$\rm{mMOM}$-like schemes,  is that  the renormalization constant of the 
gluon-ghost-antighost  vertex is 
fixed by the its equality to the renormalization constant of the same 
vertex, computed in  the ${\rm{\overline{MS}}}$-scheme:
\begin{equation}
\label{basicreq}
Z^{{\rm{mMOM}}}_{cg}(\alpha_s^{{\rm{MOM}}}, \epsilon)=
Z^{{\rm{\overline{MS}}}}_{cg}(\alpha_s^{\rm{\overline{MS}}}, \epsilon)
\end{equation}
 where 
$2\epsilon=4-D$.  This is the most important requirement 
which
allocates this scheme among all 
the variants of $\rm{MOM}$-like schemes in QCD  and greatly simplifies 
concrete calculations. Let us fix the  following   
notations for the QCD  renormalization constants:
\begin{equation}
\label{renconstants}
A^a_{B, \; \mu}=\sqrt{Z_A}A^a_{\mu}, ~~~ c^a_B=\sqrt{Z_c}c^a, ~~~ g_B=\mu^{\varepsilon}Z_gg, ~~~ \xi_B=Z^{-1}_{\xi}Z_A\xi~,
\end{equation}
where $A^a_{\mu},~ c^a$ are the renormalized gluon and ghost fields correspondingly,
 $g$ is 
the coupling constant, $\xi$ is the gauge parameter, which is included in the QCD Lagrangian in the form of additive term $(\partial_\mu A^a_\mu)^2/(2\xi)$, $\mu$ is a scale parameter of the dimensional regularization. 
As usually  the index $``B"$ denotes the bare unrenormalized quantities. One should emphasize that we work within the theory with linear covariant gauge and this fact means that we imply $Z_{\xi}=1$ in all orders of PT. The defined in Eq.(\ref{renconstants}) 
 renormalization constants are related to the renormalization 
constant of the gluon-ghost-antighost vertex as   $Z_{cg}=Z_gZ^{1/2}_AZ_c$.

In general the
 $\rm{MOM}$-like schemes are determined by the requirement that at
$Q^2=-q^2=\mu^2$
the residues for the gluon and ghost propagators are equal to unity. 
The renormalized two-point Green functions for the  gluon and ghost fields
 can be  written down in the following form:
\begin{gather}
G^{\mu\nu}_{ab}(q)=-\frac{\delta_{ab}}{q^2}\bigg[\bigg(-g^{\mu\nu}+\frac{q^\mu q^\nu}{q^2}\bigg)\frac{1}{1+\Pi_A(q^2)}-\xi\frac{q^\mu q^\nu}{q^2}\bigg]~, \\
\Delta^{ab}(q)=-\frac{\delta^{ab}}{q^2}\frac{1}{1+\Pi_c(q^2)}~,
\end{gather}
where  $\Pi_A(q^2)$ and $\Pi_c(q^2)$ are  the gluon and ghosts  
self-energy functions. 

At the subtraction point  $q^2=-\mu^2$  in the  $\rm{mMOM}$ 
scheme the requirements for the
residues of these  propagators imply the fulfillment of the following conditions:
\begin{equation}
\Pi_A^{{\rm{mMOM}}}(q^2=-\mu^2)=0~, ~~~~~ \Pi_c^{{\rm{mMOM}}}(q^2=-\mu^2)=0~.
\end{equation}
The relation between unrenormalized and renormalized two-point Green functions 
can be presented in the following form:
\begin{gather}
\hspace{-0.3cm}
\label{functionmMOM}
1+\Pi^{{\rm{mMOM}}}_A(a_s, \xi)=Z^{{\rm{mMOM}}}_A(a_s, \xi, \epsilon) \bigg(1+\Pi^B_A(a^B_s(a_s), \xi^B(\xi), \epsilon)\bigg)~, \\
\hspace{-0.2cm}
\label{Polarfuncxi}
  1+\Pi^{{\rm{mMOM}}}_c(a_s, \xi)=Z^{{\rm{mMOM}}}_c(a_s, \xi, \epsilon) \bigg(1+\Pi^B_c(a^B_s(a_s), \xi^B(\xi), \epsilon)\bigg)~,
\end{gather}
 where the QCD coupling constant  $a_s=a_s(\mu^2)$ and the gauge parameter $\xi=\xi(\mu^2)$ are  renormalized in 
the  $\rm{mMOM}$-scheme.

The similar relations hold in the $\rm{\overline{MS}}$-scheme but with a replacement of the
$\rm{mMOM}$ quantities $a_s(\mu^2),~ \xi(\mu^2)$ by their analogies, defined in the $\rm{\overline{MS}}$-scheme.

Combining the given above  definition of the renormalization constant of 
gluon-ghost-antighost vertex with 
  (\ref{basicreq}) and  (\ref{renconstants}), one gets  
the following relation between  coupling constants of the  
 $\rm{mMOM}$ and $\rm{\overline{MS}}$ schemes:
\begin{equation}
\label{Smekal}
a^{{\rm{mMOM}}}_s(\mu^2)=\frac{Z^{{\rm{mMOM}}}_A}{Z^{{\rm{\overline{MS}}}}_A}\bigg(\frac{Z^{{\rm{mMOM}}}_c}{Z^{{\rm{\overline{MS}}}}_c}\bigg)^2 a^{{\rm{\overline{MS}}}}_s(\mu^2)~,
\end{equation}
where the gauge-dependence on $\xi^{{\rm{\overline{MS}}}}$ enters in the ratios of  $Z_A$ and $Z_c$ 
in two considered schemes. One should emphasize that this relation requires knowledge of the  
renormalization constants of the gluon and ghost fields only, 
but not of any vertex structures. Using Eqs.(\ref{functionmMOM}), (\ref{Polarfuncxi}) and (\ref{Smekal}), one can obtain the following relations \cite{vonSmekal:2009ae},\cite{Gracey:2013sca},\cite{Ruijl:2017eht}:
\begin{gather}
\label{connection1}
a^{{\rm{mMOM}}}_s(\mu^2)=\frac{a^{{\rm{\overline{MS}}}}_s(\mu^2)}{\bigg(1+\Pi^{{\rm{\overline{MS}}}}_A(a^{{\rm{\overline{MS}}}}_s(\mu^2), \xi^{{\rm{\overline{MS}}}}(\mu^2))\bigg)\bigg(1+\Pi^{{\rm{\overline{MS}}}}_c(a^{{\rm{\overline{MS}}}}_s(\mu^2), \xi^{{\rm{\overline{MS}}}}(\mu^2)) \bigg)^2}~, \\
\label{connection2}
\xi^{{\rm{mMOM}}}(\mu^2)=\bigg(1+\Pi^{{\rm{\overline{MS}}}}_A(a^{{\rm{\overline{MS}}}}_s(\mu^2), \xi^{{\rm{\overline{MS}}}}(\mu^2))\bigg)\xi^{{\rm{\overline{MS}}}}(\mu^2)~.
\end{gather}
 The three-loop results for self-energies $\Pi^{{\rm{\overline{MS}}}}_A$ and $\Pi^{{\rm{\overline{MS}}}}_c$  were calculated in  Ref.\cite{Chetyrkin:2000dq} 
with explicit dependence on the 
 gauge parameter  $\xi^{{\rm{\overline{MS}}}}$
taken into account.  The analogous
four-loop results were obtained  in the recent work of Ref.\cite{Ruijl:2017eht}. Using results of these computations in the $\mathcal{O}(a^3_s)$ approximation and the  expansions (\ref{connection1}-\ref{connection2}),
 we obtain the following  relation for  $a^{\rm{\overline{MS}}}_s$ coupling constant, expressed through $a^{{\rm{mMOM}}}_s$ and ${\color{blue}\xi}=\xi^{{\rm{mMOM}}}$ (see Appendix A):
 \begin{subequations}
\begin{align}
\label{aMS-mMOM}
a^{\rm{\overline{MS}}}_s &=a^{{\rm{mMOM}}}_s +b^{{\rm{mMOM}}} _1(a^{{\rm{mMOM}}}_s)^2+b^{{\rm{mMOM}}} _2(a^{{\rm{mMOM}}}_s)^3+b^{{\rm{mMOM}}} _3(a^{{\rm{mMOM}}}_s)^4~, \\ \label{b1}
b^{{\rm{mMOM}}} _1&=\bigg[-\frac{169}{144}-\frac{1}{8}{\color{blue}\xi}-\frac{1}{16}{\color{blue}\xi^2}\bigg]{\color{magenta}
C_A}+\frac{5}{9}{\color{magenta} T_Fn_f}~, \\ \nonumber
b^{{\rm{mMOM}}} _2&=\bigg[-\frac{18941}{20736}+\frac{39}{128}\zeta_3+\bigg(\frac{889}{2304}-\frac{11}{64}\zeta_3\bigg){\color{blue}\xi}+\bigg(\frac{203}{2304}+\frac{3}{128}\zeta_3\bigg){\color{blue}\xi^2}-\frac{3}{256}{\color{blue}\xi^3}\bigg]{\color{magenta}
C_A^2} \\ \label{b2}
&+\bigg[-\frac{107}{648}+\frac{\zeta_3}{2}-\frac{5}{36}{\color{blue}\xi}-\frac{5}{72}{\color{blue}\xi^2} \bigg]{\color{magenta}
C_AT_Fn_f}+\bigg[\frac{55}{48}-\zeta_3\bigg]{\color{magenta} C_FT_Fn_f}+\frac{25}{81}{\color{magenta} T^2_Fn^2_f}~, \\ \label{b3}
b^{{\rm{mMOM}}} _3&=\bigg[-\frac{1935757}{2985984}+\frac{7495}{18432}\zeta_3+\frac{7805}{12288}\zeta_5+\bigg(\frac{4877}{36864}-\frac{611}{1536}\zeta_3+\frac{295}{1024}\zeta_5\bigg){\color{blue}\xi} \\ \nonumber
&+\bigg(\frac{17315}{110592}-\frac{47}{768}\zeta_3+\frac{175}{6144}\zeta_5\bigg){\color{blue}\xi^2}+\bigg(-\frac{233}{4608}+\frac{59}{1536}\zeta_3+\frac{5}{3072}\zeta_5\bigg){\color{blue}\xi^3} \\ \nonumber
&+\bigg(-\frac{235}{36864}-\frac{5}{6144}\zeta_3-\frac{35}{12288}\zeta_5\bigg){\color{blue}\xi^4}\bigg]{\color{magenta}C^3_A}+\bigg[-\frac{143}{288}-\frac{37}{24}\zeta_3+\frac{5}{2}\zeta_5\bigg]{\color{magenta} C^2_FT_Fn_f} \\ \nonumber
 &+\bigg[\frac{25547}{20736}+\frac{107}{144}\zeta_3-\frac{5}{4}\zeta_5+\bigg(-\frac{55}{128}+\frac{3}{8}\zeta_3\bigg){\color{blue}\xi}+\bigg(-\frac{55}{256}+\frac{3}{16}\zeta_3\bigg){\color{blue}\xi^2}
 \bigg]{\color{magenta} C_FC_AT_Fn_f} \\ \nonumber
 &+\bigg[\frac{199}{31104}-\frac{223}{384}\zeta_3-\frac{5}{6}\zeta_5+\bigg(\frac{833}{13824}+\frac{11}{144}\zeta_3\bigg){\color{blue}\xi}
+\bigg(-\frac{143}{6912}-\frac{9}{128}\zeta_3\bigg){\color{blue}\xi^2}-\frac{5}{1152}{\color{blue}\xi^3} \\ \nonumber
&+\frac{5}{2304}{\color{blue}\xi^4}
\bigg]{\color{magenta} C^2_AT_Fn_f}+   \bigg[\frac{1235}{15552}+\frac{13}{36}\zeta_3+\bigg(-\frac{19}{216}-\frac{1}{9}\zeta_3\bigg){\color{blue}\xi}
-\frac{25}{432}{\color{blue}\xi^2}\bigg]{\color{magenta} C_AT^2_Fn^2_f}
\\ \nonumber
&+\bigg[\frac{1249}{2592}-\frac{11}{18}\zeta_3\bigg]{\color{magenta} C_FT^2_Fn^2_f}+\frac{125}{729}{\color{magenta}T^3_Fn^3_f}~.
\end{align}
\end{subequations}
In the necessary order of accuracy, required for our goals, the corresponding PT relation between gauge parameters, defined in the $\rm{\overline{MS}}$ and $\rm{mMOM}$-schemes 
reads: 
\begin{subequations}
\begin{align}
\label{ksiMSAtoMOM}
\xi^{\rm{\overline{MS}}}&=\xi^{{\rm{mMOM}}}\bigg(1+\eta^{{\rm{mMOM}}} _1a^{{\rm{mMOM}}}_s+\eta^{{\rm{mMOM}}} _2(a^{{\rm{mMOM}}}_s)^2\bigg)~, \\ 
\eta^{{\rm{mMOM}}} _1&=\bigg[\frac{97}{144}+\frac{1}{8}{\color{blue}\xi}+\frac{1}{16}{\color{blue}\xi^2}\bigg]{\color{magenta} C_A}-\frac{5}{9}{\color{magenta} T_Fn_f}~, \\ 
\label{ksi2}
\eta^{{\rm{mMOM}}} _2&=\bigg[\frac{5591}{4608}-\frac{3}{16}\zeta_3+\bigg(-\frac{121}{1536}+\frac{1}{8}\zeta_3\bigg){\color{blue}\xi}+\frac{7}{256}{\color{blue}\xi^2}+\frac{7}{256}{\color{blue}\xi^3}+\frac{1}{256}{\color{blue}\xi^4}\bigg]{\color{magenta} C^2_A} \\ \nonumber
&+\bigg[-\frac{371}{576}-\frac{\zeta_3}{2}\bigg]{\color{magenta} C_AT_Fn_f}+\bigg[-\frac{55}{48}+\zeta_3\bigg]{\color{magenta} C_FT_Fn_f}~.
\end{align}
\end{subequations}
One should note that expressions similar to (\ref{aMS-mMOM}) and (\ref{ksiMSAtoMOM}) 
were presented in the works 
\cite{vonSmekal:2009ae}, \cite{Gracey:2013sca} and  \cite{Ruijl:2017eht} in terms of $\xi^{{\rm{\overline{MS}}}}$
for the  $SU(N_c)$ color  
group.  In our further analysis we will use 
the results, totally related to the $\rm{mMOM}$-scheme, transforming 
everywhere  the dependence on  the gauge parameter $\xi^{\rm{\overline{MS}}}$ 
to its analog 
in the $\rm{MOM}$-scheme.  

The ${\rm{mMOM}}$-scheme  $\beta$-function can be computed 
 with the 
explicit dependence on   $\xi^{{\rm{mMOM}}}$ using the 
following relation:
\begin{equation}
\label{bmMOM-MS}
\beta^{{\rm{mMOM}}}(a^{{\rm{mMOM}}}_s, \xi^{{\rm{mMOM}}})=\beta^{\rm{\overline{MS}}}(a_s^{{\rm{\overline{MS}}}})\frac{\partial a^{{\rm{mMOM}}}_s}{\partial a^{\rm{\overline{MS}}}_s}+\xi^{\rm{\overline{MS}}}\gamma^{\rm{\overline{MS}}}_{\xi}(a_s^{{\rm{\overline{MS}}}}, \xi^{\rm{\overline{MS}}}) \frac{\partial a^{{\rm{mMOM}}}_s}{\partial \xi^{\rm{\overline{MS}}}}~,
\end{equation}
where $\gamma_{\xi}=d\log\xi/d\log\mu^2$ is the anomalous dimension of gauge, 
which due to the Slavnov--Taylor identities coincides with the expression for 
the gluon anomalous dimension taken with the opposite sign. 
The three-loop $\rm{\overline{MS}}$-results for this anomalous dimension 
are known for an arbitrary $SU(N_c)$ color group in Ref.\cite{Larin:1993tp}. 
The corresponding expression for the RG $\beta$-function with the  arbitrary gauge in $\rm{mMOM}$-scheme,  that  can be obtained from Eq.(\ref{bmMOM-MS}), 
was presented in Ref.\cite{vonSmekal:2009ae} in the case of  $SU(N_c)$ color gauge 
group with explicit dependence on number of colors $N_c$ and number of active flavors $n_f$, and 
in Refs.\cite{Gracey:2013sca},\cite{Ruijl:2017eht} for $SU(N_c)$ group when the corresponding Casimir operators were kept non-expanded on number of colors.
As in all previously used MOM-like QCD schemes  (see e.g. 
\cite{Hagiwara:1982ct, Tarasov:1990ps})
the $\rm{mMOM}$-scheme  $\beta$-function  starts to  depend on the gauge from the  two-loop level:
\begin{subequations}
\begin{align}
\label{b0Lg}
\beta^{{\rm{mMOM}}}_0&=\frac{11}{12}{\color{magenta} C_A}-\frac{1}{3}{\color{magenta} T_Fn_f}~, \\
\label{b1Lg}
\beta^{{\rm{mMOM}}}_1&=\bigg[\frac{17}{24}-\frac{13}{192}{\color{blue}\xi}-\frac{5}{96}{\color{blue}\xi^2}+\frac{1}{64}{\color{blue}\xi^3}\bigg]{\color{magenta}C^2_A}-\frac{1}{4}{\color{magenta} C_FT_Fn_f} \\ \nonumber 
&+\bigg[-\frac{5}{12}+\frac{1}{24}{\color{blue}\xi}+\frac{1}{24}{\color{blue}\xi^2}\bigg]{\color{magenta}C_AT_Fn_f}~, \\ 
\label{b2Lg} 
\beta^{{\rm{mMOM}}}_2&=\bigg[\frac{9655}{4608}-\frac{143}{512}\zeta_3+\bigg(-\frac{1097}{6144}+\frac{33}{512}\zeta_3\bigg){\color{blue}\xi}+\bigg(-\frac{725}{6144}+\frac{13}{512}\zeta_3\bigg){\color{blue}\xi^2} \\ \nonumber
&+\bigg(\frac{21}{2048}-\frac{3}{512}\zeta_3\bigg){\color{blue}\xi^3}
+\frac{55}{6144}{\color{blue}\xi^4}\bigg]{\color{magenta} C^3_A}+\frac{1}{32}{\color{magenta}  C^2_FT_Fn_f}+ 
\bigg[\frac{23}{96}+\frac{\zeta_3}{6}\bigg]{\color{magenta}  C_AT^2_Fn^2_f} \\ \nonumber
&+\bigg[-\frac{2009}{1152}-\frac{137}{384}\zeta_3+\frac{37}{384}{\color{blue}\xi} 
+\bigg(\frac{23}{256}-\frac{\zeta_3}{128}\bigg){\color{blue}\xi^2}+\frac{1}{128}{\color{blue}\xi^3}-\frac{1}{768}{\color{blue}\xi^4}
\bigg]{\color{magenta}  C^2_AT_Fn_f} \\ \nonumber
&+\bigg[-\frac{641}{576}+\frac{11}{12}\zeta_3+\frac{1}{16}{\color{blue}\xi}+\frac{3}{64}{\color{blue}\xi^2}
\bigg]{\color{magenta}  C_AC_FT_Fn_f}  
+\bigg[\frac{23}{72}-\frac{\zeta_3}{3}\bigg]{\color{magenta}  C_FT^2_Fn^2_f}~. 
\end{align}
\end{subequations}
The two-loop coefficient $\beta^{{\rm{mMOM}}}_1$ contains cubic term $\xi^3$ and
coincides with its   $\rm{\overline{MS}}$  analogue 
at $\xi^{\rm{mMOM}}=0,\; -1$ only, 
while  the three-loop coefficient $\beta^{{\rm{mMOM}}}_2$, evaluated at $\xi=0$ and $\xi=-1$, differs from  its gauge-independent  $\rm{\overline{MS}}$ analogue, presented in Ref.\cite{Tarasov:1980au}, \cite{Larin:1993tp}. We remind here  that at fixed renormalization point
$q^2=-\mu^2$ the values of  the gauge parameter and the  coupling constant  are  fixed
and start to run when the energy scales  move away  from their initial normalization values  $\mu^2$. 

In order to study what will happen with  
 the property of the factorization of $\beta$-function in the GCR in 
the gauge-dependent $\rm{mMOM}$-scheme at the   $\mathcal{O}(a^4_s)$ level  we 
need to  analyze the 
product of the analytical  
general $SU(N_c)$ expressions 
for the non-singlet contributions to the Adler function 
and Bjorken polarized sum rule in this gauge-dependent scheme. These expressions will be obtained below. 

\subsection{The NS Adler function in the $\rm{mMOM}$-scheme in the  $\mathcal{O}(a^4_s)$  approximation}
Using the $\mathcal{O}(a^4_s)$ result for the 
$D^{NS}(a_s)$-function in the $\rm{\overline{MS}}$-scheme, 
presented in Ref.\cite{Baikov:2010je}, and the  results of 
Eqs.(\ref{aMS-mMOM}-\ref{ksi2}), 
  we obtain the following analytical expression for
this Eucledian  characteristic of the $e^+e^-\rightarrow \gamma \rightarrow  {hadrons}$ process in the $\rm{mMOM}$-scheme:  
  \vspace{-0.4cm}
  \begin{subequations}
\begin{align}
\label{DmMOMadl}
D^{NS}_{{\rm{mMOM}}}&=1+\sum\limits_{k=1}^4 d^{{\rm{mMOM}}}_k(a^{{\rm{mMOM}}}_s)^k~, \\
\label{d1mMOM}
d^{{\rm{mMOM}}}_1&=\frac{3}{4}{\color{magenta} C_F}~, \\ 
\label{d2mMOM}
d^{{\rm{mMOM}}}_2&=-\frac{3}{32}{\color{magenta}C^2_F}+
\bigg[\frac{569}{192}-\frac{11}{4}\zeta_3-\frac{3}{32}{\color{blue}\xi}-\frac{3}{64}{\color{blue}\xi^2}\bigg]{\color{magenta} C_FC_A}+\bigg[-\frac{23}{24}+\zeta_3\bigg]{\color{magenta} C_FT_Fn_f} \\
\label{d3mMOM}
d^{{\rm{mMOM}}}_3&=-\frac{69}{128}{\color{magenta}C^3_F}+
\bigg[-\frac{1355}{768}-\frac{143}{16}\zeta_3+\frac{55}{4}\zeta_5+\frac{3}{128}{\color{blue}\xi}+\frac{3}{256}{\color{blue}\xi^2} \bigg]{\color{magenta}C^2_FC_A} \\ \nonumber
&+
\bigg[-\frac{2033}{192}+\frac{89}{12}\zeta_3+\frac{5}{6}\zeta_5+\bigg(\frac{23}{96}-\frac{\zeta_3}{4}\bigg){\color{blue}\xi}+\bigg(\frac{23}{192}-\frac{\zeta_3}{8}\bigg){\color{blue}\xi^2}\bigg] {\color{magenta}C_FC_AT_Fn_f} \\ \nonumber
&+\bigg[\frac{29}{96}+4\zeta_3-5\zeta_5\bigg]{\color{magenta}C^2_FT_Fn_f}+ 
\bigg[\frac{3}{2}-\zeta_3\bigg]{\color{magenta} C_FT^2_Fn^2_f}+\bigg[\frac{50575}{3072}-\frac{18929}{1536}\zeta_3 
-\frac{55}{24}\zeta_5 \\ \nonumber
&+\bigg(-\frac{2063}{3072}+\frac{143}{256}\zeta_3\bigg){\color{blue}\xi}+\bigg(-\frac{1273}{3072}+\frac{185}{512}\zeta_3\bigg){\color{blue}\xi^2}-\frac{9}{1024}{\color{blue}\xi^3}\bigg]{\color{magenta} C_FC^2_A} 
~, \\
\label{d4mMOM}
d^{{\rm{mMOM}}}_4&={\color{magenta}\frac{d^{abcd}_Fd^{abcd}_A}{N_c}}\bigg[\frac{3}{16}-\frac{\zeta_3}{4}-\frac{5}{4}\zeta_5\bigg]+{\color{magenta} n_f\frac{d^{abcd}_Fd^{abcd}_F}{N_c}}\bigg[-\frac{13}{16}-\zeta_3+\frac{5}{2}\zeta_5\bigg] \\ \nonumber
&+\bigg[-\frac{12305}{2048}-\frac{139}{128}\zeta_3+\frac{2255}{32}\zeta_5-\frac{1155}{16}\zeta_7+\frac{207}{1024}{\color{blue}\xi}+\frac{207}{2048}{\color{blue}\xi^2}\bigg]{\color{magenta} C^3_FC_A} \\ \nonumber
&+\bigg[-\frac{1850345}{73728}-\frac{509815}{6144}\zeta_3+\frac{5575}{64}\zeta_5+\frac{1155}{32}\zeta_7+\bigg(\frac{2639}{4096}+\frac{3465}{1024}\zeta_3-\frac{165}{32}\zeta_5\bigg){\color{blue}\xi} \\ \nonumber
&+\bigg(\frac{697}{2048}+\frac{3423}{2048}\zeta_3-\frac{165}{64}\zeta_5\bigg){\color{blue}\xi^2}+\frac{3}{4096}{\color{blue}\xi^3}-\frac{3}{8192}{\color{blue}\xi^4}\bigg]{\color{magenta}C^2_FC^2_A}+ 
\bigg[\frac{4157}{2048}+\frac{3}{8}\zeta_3\bigg]{\color{magenta} C^4_F}  \\ \nonumber
&+
\bigg[\frac{5674729}{49152}-\frac{1504181}{24576}\zeta_3 
-\frac{2907785}{49152}\zeta_5-\frac{385}{64}\zeta_7+ 
\frac{4411}{256}\zeta^2_3 +\bigg(-\frac{828805}{147456} \\ \nonumber
&+\frac{15845}{6144}\zeta_3
+\frac{4405}{4096}\zeta_5+\frac{121}{128}\zeta^2_3\bigg){\color{blue}\xi}+\bigg(-\frac{171455}{49152}+\frac{5667}{2048}\zeta_3+\frac{3695}{8192}\zeta_5-\frac{33}{256}\zeta^2_3\bigg){\color{blue}\xi^2}
\\ \nonumber
&+\bigg(-\frac{835}{12288}+\frac{103}{2048}\zeta_3+\frac{5}{4096}\zeta_5\bigg){\color{blue}\xi^3} +\bigg(\frac{503}{49152}-\frac{93}{8192}\zeta_3-\frac{35}{16384}\zeta_5\bigg){\color{blue}\xi^4}
 \bigg] {\color{magenta}C_FC^3_A} 
\\ \nonumber
&+\bigg[\frac{287}{256}+\frac{17}{8}\zeta_3-\frac{235}{8}\zeta_5+\frac{105}{4}\zeta_7\bigg]{\color{magenta} C^3_FT_Fn_f} +\bigg[-\frac{67}{24}+\frac{7}{6}\zeta_3+\frac{5}{3}\zeta_5\bigg]{\color{magenta} C_FT^3_Fn^3_f}\\ \nonumber
&+ \bigg[\frac{48451}{4608}+\frac{2109}{32}\zeta_3-\frac{1085}{16}\zeta_5-\frac{105}{8}\zeta_7 
-\frac{11}{4}\zeta^2_3+\bigg(-\frac{29}{256}-\frac{3}{2}\zeta_3+\frac{15}{8}\zeta_5\bigg){\color{blue}\xi} \\ \nonumber
&+\bigg(-\frac{29}{512}-\frac{3}{4}\zeta_3+\frac{15}{16}\zeta_5\bigg){\color{blue}\xi^2}\bigg]{\color{magenta}C^2_FC_AT_Fn_f}
+\bigg[-\frac{2027833}{18432}+\frac{12977}{256}\zeta_3+\frac{695}{12}\zeta_5\\ \nonumber
&+\frac{35}{16}\zeta_7-\frac{489}{64}\zeta^2_3
+\bigg(\frac{68479}{18432}-\frac{493}{256}\zeta_3-\frac{5}{16}\zeta_5-\frac{11}{32}\zeta^2_3\bigg){\color{blue}\xi} 
+ \bigg(\frac{6409}{3072}-\frac{197}{128}\zeta_3-\frac{5}{32}\zeta_5 \\ \nonumber
&+\frac{3}{64}\zeta^2_3\bigg){\color{blue}\xi^2}
+\bigg(\frac{23}{3072}-\frac{1}{128}\zeta_3\bigg){\color{blue}\xi^3} 
+\bigg(-\frac{23}{6144}+\frac{1}{256}\zeta_3\bigg){\color{blue}\xi^4}
\bigg]{\color{magenta} C_FC^2_AT_Fn_f} \\ \nonumber
&+\bigg[\frac{73339}{2304}-\frac{429}{32}\zeta_3-\frac{35}{2}\zeta_5+\frac{\zeta^2_3}{2}+\bigg(-\frac{13}{24}+\frac{7}{24}\zeta_3\bigg){\color{blue}\xi} \\ \nonumber
&+\bigg(-\frac{9}{32}+\frac{3}{16}\zeta_3\bigg){\color{blue}\xi^2}
\bigg]{\color{magenta} C_FC_AT^2_Fn^2_f} 
+\bigg[-\frac{125}{384}-\frac{281}{24}\zeta_3+\frac{25}{2}\zeta_5+\zeta^2_3\bigg]{\color{magenta} C^2_FT^2_Fn^2_f}
~.
\end{align}
\end{subequations}
Note that in $SU(N_c)$  theory  we have: 
\begin{gather*} 
C_F=\frac{N^2_c-1}{2N_c}~, ~~ C_A=N_c~, ~~T_F=\frac{1}{2}~, \\
\frac{d_F^{abcd}d_A^{abcd}}{N_c}=\frac{(N^2_c-1)(N^2_c+6)}{48}~, ~~\frac{d_F^{abcd}d_F^{abcd}}{N_c}=\frac{(N^2_c-1)(N^4_c-6N^2_c+18)}{96N^3_c}~,
\end{gather*} 
while in  the $SU_c(3)$ case  $C_F=4/3$, $C_A=3$, $d_F^{abcd}d_A^{abcd}/N_c=5/2$, $d_F^{abcd}d_F^{abcd}/N_c=5/36$.

In the particular case of the Landau gauge $\xi=0$  with $n_f$=3 
the numerical expressions for the coefficients (\ref{d1mMOM}-\ref{d4mMOM}) 
are consistent with the ones, obtained in  Ref.\cite{vonSmekal:2009ae}. 
The detailed study of the numerical behavior of these PT series for 
other $n_f$ and other values of gauges  will be presented below.  
The scheme-dependence
of the PT series for the $R$-ratio, defined in Eq.(\ref{R}) in Minkowski region and related to Euclidean Adler function, was analyzed in the  
$\rm{mMOM}$-scheme at $\xi=0$  in  Refs.\cite{Gracey:2014pba} and   
\cite{Kataev:2015yha}.

\subsection{The NS Bjorken function in the $\rm{mMOM}$-scheme in the  $\mathcal{O}(a^4_s)$  approximation}
Let us now turn to the calculation of the $C_{Bjp}^{NS}(a_s)$-function in the  $\mathcal{O}(a^4_s)$-approximation
 in the $\rm{mMOM}$-scheme in the $SU(N_c)$ version of QCD. Its analytical expression can be obtained 
from the summarized in Ref.\cite{Baikov:2010je} $\rm{\overline{MS}}$-scheme 
results  and 
the presented  in  Sec.3.1 transformations  of the QCD 
coupling constant $a_s^{\rm{\overline{MS}}}$ from  the  $\rm{\overline{MS}}$-scheme to the gauge-dependent 
$\rm{mMOM}$-scheme with the gauge parameter $\xi$, defined in this renormalization scheme.

The corresponding expression for the  PT coefficient function of the polarized Bjorken sum rule, calculated in the $\rm{mMOM}$-scheme,
reads
  \begin{subequations}
\begin{equation}
\label{CmMOMBjr}
C^{NS}_{Bjp,\;{\rm{mMOM}}}=1+\sum\limits_{k=1}^4 c^{{\rm{mMOM}}}_k(a^{{\rm{mMOM}}}_s)^k~, \\ 
\end{equation} 
 Its coefficients $c_k$ have the  following   form  
\begin{align} 
\label{c1mMoM}
c^{{\rm{mMOM}}}_1&=-\frac{3}{4}{\color{magenta} C_F}~, \\ 
\label{c2mMOM}
c^{{\rm{mMOM}}}_2&=\frac{21}{32}{\color{magenta} C^2_F}+\bigg(-\frac{107}{192}+\frac{3}{32}{\color{blue}\xi}+\frac{3}{64}{\color{blue}\xi^2}\bigg){\color{magenta} C_FC_A}+\frac{1}{12}{\color{magenta} C_FT_Fn_f}~, \\
\label{c3mMOM}
c^{{\rm{mMOM}}}_3&=-\frac{3}{128}{\color{magenta}C^3_F}+
\bigg[\frac{1415}{2304}-\frac{11}{12}\zeta_3 -\frac{21}{128}{\color{blue}\xi}-\frac{21}{256}{\color{blue}\xi^2}\bigg]{\color{magenta}C^2_FC_A} \\ \nonumber
&+\bigg[-\frac{13}{36}+\frac{\zeta_3}{3}\bigg]{\color{magenta} C^2_FT_Fn_f}
+\bigg[\frac{13}{9}+\frac{3}{8}\zeta_3-\frac{5}{6}\zeta_5-\frac{1}{48}{\color{blue}\xi}-\frac{1}{96}{\color{blue}\xi^2}
\bigg]{\color{magenta} C_FC_AT_Fn_f} \\ \nonumber
&+\bigg[-\frac{20585}{9216}
-\frac{117}{512}\zeta_3
+\frac{55}{24}\zeta_5 
+\bigg(\frac{215}{3072}+\frac{33}{256}\zeta_3\bigg){\color{blue}\xi}+\bigg(\frac{349}{3072}-\frac{9}{512}\zeta_3\bigg){\color{blue}\xi^2} \\ \nonumber
&+\frac{9}{1024}{\color{blue}\xi^3}\bigg] {\color{magenta} C_FC^2_A}-\frac{5}{24}{\color{magenta} C_FT^2_Fn^2_f}~,
\end{align}
\begin{align}
\label{c4mMOMBj}
c^{{\rm{mMOM}}}_4&={\color{magenta} \frac{d^{abcd}_Fd^{abcd}_A}{N_c}}\bigg[-\frac{3}{16}+\frac{\zeta_3}{4}+\frac{5}{4}\zeta_5\bigg]+{\color{magenta} n_f\frac{d^{abcd}_Fd^{abcd}_F}{N_c}}\bigg[\frac{13}{16}+\zeta_3-\frac{5}{2}\zeta_5\bigg] \\ \nonumber
&+\bigg[-\frac{4823}{2048}-\frac{3}{8}\zeta_3\bigg]{\color{magenta} C^4_F}
+\bigg[-\frac{13307}{18432}-\frac{971}{96}\zeta_3+\frac{1045}{48}\zeta_5+\frac{9}{1024}{\color{blue}\xi}+\frac{9}{2048}{\color{blue}\xi^2}
\bigg]{\color{magenta} C^3_FC_A} \\ \nonumber
&+\bigg[\frac{2543485}{221184}+\frac{90169}{6144}\zeta_3-\frac{1375}{144}\zeta_5-\frac{385}{16}\zeta_7+\bigg(-\frac{1339}{12288}+\frac{121}{1024}\zeta_3\bigg){\color{blue}\xi} \\ \nonumber
&+\bigg(-\frac{1117}{6144}+\frac{415}{2048}\zeta_3\bigg){\color{blue}\xi^2}-\frac{21}{4096}{\color{blue}\xi^3}+\frac{21}{8192}{\color{blue}\xi^4}
\bigg]{\color{magenta} C^2_FC^2_A}
+\bigg[-\frac{3927799}{442368}+\frac{49763}{73728}\zeta_3 \\ \nonumber
&+\frac{345755}{147456}\zeta_5+\frac{385}{64}\zeta_7-\frac{121}{96}\zeta^2_3+\bigg(\frac{107569}{147456}+\frac{1623}{2048}\zeta_3-\frac{4405}{4096}\zeta_5\bigg){\color{blue}\xi} \\ \nonumber
&+\bigg(\frac{28303}{49152}-\frac{11}{512}\zeta_3-\frac{3695}{8192}\zeta_5\bigg){\color{blue}\xi^2}+\bigg(\frac{151}{3072}-\frac{59}{2048}\zeta_3-\frac{5}{4096}\zeta_5\bigg){\color{blue}\xi^3} \\ \nonumber
&+\bigg(-\frac{41}{49152}+\frac{5}{8192}\zeta_3+\frac{35}{16384}\zeta_5\bigg){\color{blue}\xi^4}
\bigg]{\color{magenta} C_FC^3_A}+
\bigg[\frac{317}{144}+\frac{109}{24}\zeta_3-\frac{95}{12}\zeta_5\bigg]{\color{magenta}C^3_FT_Fn_f}\\ \nonumber
&+\bigg[-\frac{6229}{864}-\frac{1739}{288}\zeta_3+\frac{205}{72}\zeta_5+\frac{35}{4}\zeta_7 
+\bigg(\frac{13}{96}-\frac{\zeta_3}{8}\bigg){\color{blue}\xi}+\bigg(\frac{13}{192}-\frac{\zeta_3}{16}\bigg){\color{blue}\xi^2}
\bigg]{\color{magenta} C^2_FC_AT_Fn_f} \\ \nonumber
&+
\bigg[\frac{12265}{1728}-\frac{1237}{512}\zeta_3+\frac{15}{16}\zeta_5 
-\frac{35}{16}\zeta_7
+\frac{11}{12}\zeta^2_3+\bigg(-\frac{8257}{18432}-\frac{49}{96}\zeta_3+\frac{5}{16}\zeta_5\bigg){\color{blue}\xi}\\ \nonumber
&+\bigg(-\frac{869}{3072}-\frac{33}{512}\zeta_3+\frac{5}{32}\zeta_5\bigg){\color{blue}\xi^2} 
-\frac{1}{1536}{\color{blue}\xi^3}+\frac{1}{3072}{\color{blue}\xi^4}
\bigg]{\color{magenta} C_FC^2_AT_Fn_f} \\ \nonumber
&+\bigg[-\frac{1283}{864}+\frac{85}{72}\zeta_3-\frac{35}{36}\zeta_5-\frac{\zeta^2_3}{6}
+\bigg(\frac{11}{192}+\frac{\zeta_3}{12}\bigg){\color{blue}\xi}+\frac{5}{128}{\color{blue}\xi^2}
\bigg]{\color{magenta} C_FC_AT^2_Fn^2_f} \\ \nonumber
&+
\bigg[\frac{1891}{3456}-\frac{\zeta_3}{36}\bigg]{\color{magenta} C^2_FT^2_Fn^2_f}  +  \frac{5}{72}  {\color{magenta} C_FT^3_Fn^3_f}~.
\end{align}
\end{subequations}
We  will use the analytical  results, presented  in Sec.3.1, 3.2  and 3.3,   for two purposes. The first problem we are interested in is 
whether  the unexplained yet  from the first principles of gauge quantum field 
theory    
$\rm{\overline{MS}}$-scheme structure of the GCR,  discovered in 
Ref.\cite{Broadhurst:1993ru},  will be 
essentially  modified in the $\rm{mMOM}$-scheme. The second problem is more    
landed and is related to the comparison  of the asymptotic  behavior of 
the $\mathcal{O}(a^4_s)$-approximations of the PT series for $D^{NS}(a_s)$ 
and $C_{Bjp}^{NS}(a_s)$-functions in the  $\rm{\overline{MS}}$-scheme and  
the  $\rm{mMOM}$-scheme for different numbers of quark flavors $n_f$ and in  
several fixed gauges. 

\subsection{About the gauge-dependence of 
the generalized Crewther relation in the $\rm{mMOM}$-scheme} 
Consider now  the question whether (or when)  the RG gauge-dependent  
 $\beta^{{\rm{mMOM}}}$-function is  
 factorized in the 
 GCR. This problem will be analyzed  using the analytical results of  
  Eqs.(\ref{b0Lg}--\ref{c4mMOMBj}). 
Taking first  into account the  relation  
(\ref{bK1}) we conclude that 
 in the $\mathcal{O}(a^2_s)$ approximation  the conformal symmetry breaking term $\Delta_{csb}$ can be factorized at any value of the gauge parameter $\xi$ thanks to the the fulfillment of Eq.(\ref{d1c1}), which follows from the
property of scheme-independence of this relation. This implies that 
 the coefficient $K_1$ does not depend on $\xi$ and coincides with its  $\rm{\overline{MS}}$ expression, namely 
\begin{equation}
\label{K1mMOM}
K^{{\rm{mMOM}}}_1=\bigg(-\frac{21}{8}+3\zeta_3\bigg)C_F~.  
\end{equation}
Using now  Eq.(\ref{bK2}) we find that  
the $\mathcal{O}(a^3_s)$ $\rm{mMOM}$-scheme  coefficient in $\Delta_{csb}$ can not be  represented in the form (\ref{BK}) for any $\xi$. 
Indeed,  equation (\ref{bK2}) imposes certain  restrictions on the factorization conditions
in $\rm{mMOM}$-scheme. Taking into account the concrete analytical 
results for the coefficients $d_k^{\rm{mMOM}}$  from Sec.3.2 
and for the coefficients $c_k^{\rm{mMOM}}$ from Sec.3.3 with $1\leq k\leq 3$ we 
find that the two-loop  $\beta^{{\rm{mMOM}}}$-function  is factorized 
in  the GCR only for certain values of gauge parameter, which are fixed 
by the solution of the following equation:
\begin{equation}
\label{EQxi}
\xi^3+4\xi^2+3\xi=0~, 
\end{equation}
 namely 
for three specific values  $\xi=0,\;-1,\;-3$. The extra more 
detailed theoretical clarification of these foundations  will be presented 
below in the separate Section.  

It should be stressed that 
the Landau gauge $\xi=0$ is often  used in multiloop calculations 
(see e.g.  \cite{vonSmekal:2009ae},\cite{Gracey:2013sca,Gracey:2014pba},\cite{Kataev:2015yha},\cite{Ruijl:2017eht}).
Its 
most vivid feature is the validity of the property of the 
non-renormalization of the gluon-ghost-antighost vertex 
\cite{Taylor:1971ff}.  That is why the renormalization constant $Z_{cg}$ in 
the  $\rm{mMOM}$-scheme is chosen the same as in the $\rm{\overline{MS}}$-scheme. 
Note also that in the Landau gauge the  longitudinal part 
of the renormalized gluon propagator vanishes and therefore its PT approximation has   a transverse structure, namely $q_{\mu}G^{\mu\nu}_{ab}(q)=0$.

Other two  gauges $\xi=-1$ and $\xi=-3$ are  less studied.  However,   some 
attractive  features  of using in the PT   QCD 
expansions of  measurable physical quantities   
the class of  MOM-schemes  with  anti-Feynman gauge $\xi=-1$ (more  
precise of  the class of 
gauges with  $\vert \xi\vert\leq 1$) was noticed in the work of \cite{Braaten:1981dv}, where it was shown that  for these values of gauge the one-loop QCD  corrections to MOM-scheme  effective charges, 
defined as the combination of Green functions,   
are  rather small at
$n_f=4$.   The anti-Yennie gauge $\xi=-3$ was first used in QCD  
by Stefanis  \cite{Stefanis:1983ke, Stefanis:2012if}  to clarify the special features of 
renormalizations of gauge-invariant definition of 
QCD quark correlator, formulated with the help of Wilson line.  
This gauge was independently applied later on by Mikhailov 
 in Refs.\cite{Mikhailov:1998xi, Mikhailov:1999ht}, where it was    demonstrated 
that when this gauge is chosen 
the one-loop correction to the renormalization constant of the 
gluon field is proportional to the first  scheme-independent coefficient   
of  the QCD  
$\beta$-function.  In what follows we   call this anti-Yennie  gauge 
$\xi=-3$ as the   Stefanis--Mikhailov gauge.

Taking now into account  Eq.(\ref{bK2}) and the analytical $\rm{mMOM}$ 
results of (\ref{b0Lg}-\ref{c4mMOMBj}), specified to the cases of the 
Landau,  anti-Feynman and Stefanis--Mikhailov gauges,  we obtain the 
explicit analytical expressions of 
the $K^{\rm{mMOM}}_2$ coefficient, included in Eq.(\ref{function}), for these 
three  gauges, 
which do not violate the 
property of  the factorization of the two-loop  $\rm{mMOM}$  $\beta$-function in the CSB term of Eq.(\ref{BK}) in the 
$\mathcal{O}(a_s^3)$ approximation of the GCR. The concrete results and 
the analysis of their structure will be presented in Section 4.

The dependence on $\xi$ of the  $\mathcal{O}(a^4_s)$ coefficient of the GCR is 
  more complicated.   We found  that for the case of the Landau 
gauge the fundamental property of factorization of the  three-loop $\rm{mMOM}$-scheme 
$\beta$-function of $SU(N_c)$ theory in the conformal symmetry breaking 
term  $\Delta_{csb}$ of the 
$\rm{mMOM}$-variant of the GCR takes place. The corresponding  analytical expression of the $K^{\rm{mMOM}}_3$-term at $\xi=0$  will be also presented in Sec.4. 

In the case of $\xi=-1$ and $\xi=-3$  only \textit{partial factorization} of three-loop $\rm{mMOM}$ $\beta$-function in 
the analytical expression for the CSB term of the GCR is  observed. Indeed,   
the  analytical expression of the $K_3$-term  contains six color structures, first revealed in the 
case of application of the  gauge-independent 
$\rm{\overline{MS}}$-scheme in  Ref.\cite{Baikov:2010je}. In  the   $\rm{mMOM}$-scheme when the Landau 
gauge is chosen,  $K_3$  also contains these six color structures and we 
conclude that in this case total factorization of the three-loop QCD $\beta$-function also persists 
in the GCR.  However, in the case of anti-Feynman and Stefanis--Mikhailov gauges only
five from these six  color structures may be found from Eq.(\ref{bK3}) and the one 
is not determined, namely the coefficient, proportional to  $C_FC_AT_Fn_f$-contribution (see Appendix B) 
Therefore   
for these two gauges  the three-loop approximation of the 
$\rm{mMOM}$-scheme $\beta$-function obeys the property of  \textit{partial factorization} only.

Thus from the study of the structure of the GCR in the $\rm{mMOM}$-scheme in the Landau gauge 
we come to  conclusion that  the property of \textit{gauge invariance} of the renormalization schemes 
\textit{is the sufficient but not a necessary} property  for the factorization of the  $SU(N_c)$ $\beta$-function 
in the $\mathcal{O}(a_s^4)$-expression of the   generalized Crewther relation in QCD, 
presented in the concrete  renormalization schemes, which are fixed by kinematics conditions of the subtractions of renormalizations 
of the Green functions, related to the QCD vertexes. 

\subsection{Asymptotic behavior  of the NS Adler and Bjorken functions at the fourth-loop level:   the 
$\rm{mMOM}$ vs   $\rm{\overline{MS}}$-scheme}

Let us study the asymptotic behavior of the  $\mathcal{O}(a_s^4)$-approximation for 
two basic functions $D^{NS}(a_s)$ and $C^{NS}_{Bjp}(a_s)$, which enter the GCR and  both were 
extracted 
from the concrete  experimental data (see \cite{Eidelman:1998vc, Adolph:2016myg}). From phenomenological point of view one may 
realize that 
these studies are  of interest for  the  values  $n_f=3,4,5$, while the results, 
summarized above may attract definite interest to the behavior of the  corresponding
PT series, obtained within $\rm{mMOM}$ scheme with   three specific values  $\xi=0,\;-1,\;-3$ of 
the gauge parameter.

These results are  compared with the $\rm{\overline{MS}}$-results in the   $\mathcal{O}(a_s^4)$-approximation 
and are presented in the Table 1. Note, that the
 $\rm{mMOM}$-scheme  numerical results for $D^{NS}(a_s)$ 
agree with the ones, obtained in Ref.\cite{vonSmekal:2009ae} for the case of the Landau gauge. It was also recently checked that  
the given in  the
Table 1 $\mathcal{O}(a_s^4)$  $\rm{mMOM}$ expressions  for $C^{NS}_{Bjp}$ at $\xi=0$, 
  presented in Ref.\cite{Kataev:2017oqg}, are consistent with the   results of
 Ref.\cite{Ayala:2017uzx}, based  on the QCD calculation  within  
the variant of Analytical Perturbation Theory (APT), developed previously  in Ref.\cite{Shirkov:2000qv}\footnote{We are grateful to G. Cveti\v c for the confirmation    
of this agreement.}. 
\begin{table}[h!]
\hspace{-1.5cm}
\begin{floatrow}[1]
{
{\def\arraystretch{1.3}\tabcolsep=0.1pt
\begin{tabular}{|c|c|c|c|}
\hline 
$\xi$ & $n_f$  & $
\makecell{ \\ 
\textbf{$\;\;\;$The flavor NS Adler function} ~ \bf{D^{NS}} ~~~ \\ 
\textbf{in} ~   \bf{{\rm{mMOM}}}~        
\textbf{and} ~ \bf{{{\rm{\overline{MS}}}}}~ \textbf{schemes} }$ & $\makecell{ \\
\textbf{$\;\;\;$The flavor NS Bjorken function} ~ \bf{C^{NS}_{Bjp}}~~~ \\ \textbf{in} ~    \bf{{\rm{mMOM}}}~  \textbf{and} ~  \bf{{{\rm{\overline{MS}}}}}~
\textbf{schemes}  }$ \\ \hline
\cline{2-4}
\multirow{3}{*}{$\;$ 0 $\;$} & $\;$ 3 $\;$ &  $1+a_s-1.048a^2_s-4.8241a^3_s+3.12575a^4_s$ & $1-a_s-0.896a^2_s+1.4262a^3_s-22.96225a^4_s$ \\ \cline{2-4}
& 4 & $1+a_s-0.885a^2_s-5.8133a^3_s+11.71854a^4_s$ & $1-a_s-0.840a^2_s+3.0375a^3_s-12.34185a^4_s$ \\
\cline{2-4}
& 5 & $1+a_s-0.723a^2_s-6.6039a^3_s+18.25793a^4_s$ & $1-a_s-0.785a^2_s+4.5099a^3_s-3.61660a^4_s$ \\
\hline
\cline{2-4}
\multirow{3}{*}{-1} & 3 & $1+a_s-0.860a^2_s-4.3579a^3_s+6.38863a^4_s$ & $1-a_s-1.083a^2_s+0.2312a^3_s-31.54404a^4_s$ \\
\cline{2-4}
& 4 & $1+a_s-0.698a^2_s-5.2862a^3_s+13.28190a^4_s$ & $1-a_s-1.028a^2_s+1.8633a^3_s-18.49192a^4_s$ \\
\cline{2-4}
& 5 & $1+a_s-0.535a^2_s-6.0159a^3_s+18.39217a^4_s$ & $1-a_s-0.972a^2_s+3.3566a^3_s-7.57175a^4_s$ \\
\hline
\cline{2-4}
\multirow{3}{*}{-3} & 3 & $1+a_s-1.610a^2_s-0.1797a^3_s+15.90258a^4_s$ & $1-a_s-0.333a^2_s-1.0317a^3_s-44.09174a^4_s$  \\
\cline{2-4}
& 4 & $1+a_s-1.448a^2_s-1.3517a^3_s+23.10284a^4_s$ & $1-a_s-0.278a^2_s+0.5170a^3_s-31.54819a^4_s$  \\
\cline{2-4}
& 5 & $1+a_s-1.285a^2_s-2.3251a^3_s+28.39054a^4_s$ & $1-a_s-0.222a^2_s+1.9269a^3_s-21.14145a^4_s$ \\
\hline
\cline{2-4}
\multirow{3}{*}{---} & 3 & $1+\bar{a}_s+1.640\bar{a}^2_s+6.3710\bar{a}^3_s+49.07570\bar{a}^4_s$ & $1-\bar{a}_s-3.583\bar{a}^2_s-20.2153\bar{a}^3_s-175.74950\bar{a}^4_s$ \\
\cline{2-4}
& 4 & $1+\bar{a}_s+1.525\bar{a}^2_s+2.7586\bar{a}^3_s+27.38880\bar{a}^4_s$ & $1-\bar{a}_s-3.250\bar{a}^2_s-13.8503\bar{a}^3_s-102.40202\bar{a}^4_s$  \\
\cline{2-4}
& 5 & $1+\bar{a}_s+1.409\bar{a}^2_s-0.6814\bar{a}^3_s+9.21018\bar{a}^4_s$ & $1-\bar{a}_s-2.917\bar{a}^2_s-7.8402\bar{a}^3_s-41.95977\bar{a}^4_s$  \\
\hline
\end{tabular}}}
\vspace{0.2cm}
{\caption*{ TABLE 1. 
The comparison of the $\mathcal{O}(a^4_s)$   PT expansions  for the  
$D^{NS}$ and $C^{NS}_{Bjp}$ functions, evaluated  
in  QCD  with $n_f=3,4,5$ active flavors  in   
$\rm{mMOM}$-scheme  with   $\xi=0, -1, -3$ 
and $\rm{\overline{MS}}$-scheme ($\bar{a}_s$ corresponds to the calculation in the $\rm{\overline{MS}}$-scheme).}}
\end{floatrow}
\end{table}
The existing at present point of view on the structure of the 
asymptotic 
PT QCD  series for the physical quantities, defined in the Euclidean region,
indicates  
that in the $\rm{\overline{MS}}$-scheme the coefficients of these 
PT expansions  obey the pattern 
of infrared  renormalon (IRR)  generated    
 sign-constant factorial growth up to the level, when they are 
starting to compete with sign-alternating  factorial contributions,  
generated in the related  PT QCD  series by the 
corresponding  ultraviolet renormalon (UVR) effects 
(for  the detailed discussions see reviews \cite{Beneke:1998ui, Beneke:2000kc}). 
The concrete $\rm{\overline{MS}}$-scheme  
calculations of the Borel image for the $D^{NS}(a_s)$-function 
\cite{Beneke:1992ch}  and the one  
for the $C^{NS}_{Bjp}(a_s)$-function  \cite{Broadhurst:1993ru} demonstrate
 the  interplay 
of the  IRR and UVR effects are manifesting itself in the case of 
$D^{NS}(a_s)$ at 
the level  above $\mathcal{O}(a_s^4)$ contribution and for the $C^{NS}_{Bjp}(a_s)$ 
even at more high  level of the related  PT expansions  \cite{Beneke:1998ui}.

The comparison of numerical $\rm{\overline{MS}}$-results, presented in 
Table 1,  give extra argument in flavor of this 
renormalon-motivated guess. Indeed, in the case of the Bjorken polarized sum 
rule the the sign-constant pattern   and the related asymptotic growth 
of the coefficients of $\mathcal{O}(a_s^4)$-approximation 
for all considered values of $n_f=3,4,5$ are more pronounced, than in the 
case of $D^{NS}(a_s)$-function. Moreover, for  $n_f=5$ in the latter case  these features are even 
violated at the $\mathcal{O}(a_s^3)$ approximation.     
 The inconstancy of signs of PT series in the $\rm{mMOM}$-scheme is observed for both physical quantities 
(apart of the case of $\rm{mMOM}$-PT series for $C_{Bjp}^{NS}(a_s)$ with $n_f=3$ and 
$\xi=-3$). However, without any estimates of the numerical values 
of the unknown at present $\mathcal{O}(a_s^5)$ corrections we can not make definite 
conclusion whether in the $\rm{mMOM}$-scheme the considered  asymptotic  PT  QCD series have   
sign-alternating or sign-nonregular structure. Next, on the contrary to the $\rm{\overline{MS}}$ PT approximations for 
$D^{NS}(a_s)$-function  of  their    
$\rm{mMOM}$-scheme analogs are growing when $n_f$ is increasing from $n_f=3$ to $n_f=5$. 
The situation for the Bjorken polarized sum rule  function is somewhat different: here with growth of $n_f$   
the values of the $\mathcal{O}(a^4_s)$ coefficients are decreasing  modulo in both schemes. One more interesting 
feature of the $\rm{mMOM}$-scheme  PT series  for  $C^{NS}_{Bjp}$  catches the eyes: the absolute values of
the  $a_s^2$, $a_s^3$ and $a_s^4$ coefficients   are much  smaller than the ones, 
obtained in $\rm{\overline{MS}}$-scheme. This difference may be essential in the process 
of study of the scheme-dependence uncertainties of the possible new more careful  analysis of the experimental 
data for the Bjorken polarized sum rule (see Ref.\cite{Ayala:2017uzx}), which  should include   
virtual heavy-quark massive  effects, calculated at the leading order of PT  in Ref.\cite{Teryaev:1996nf}
and next-to-leading order in Ref.\cite{Blumlein:1998sh} and in the most  detailed  work of Ref.\cite{Blumlein:2016xcy}.

\section{The generalized Crewther relation in the MOM-like schemes in QCD} 

\subsection{The $\mathcal{O}(a_s^4)$ $\beta$-factorization of the generalized Crewther relation in the  $\rm{mMOM}$-scheme}  

Let us  return to analysis of the analytical 
structure of the $\mathcal{O}(a_s^4)$ approximation of the GCR in $SU(N_c)$  QCD in the case of application of the gauge-dependent 
${\rm{mMOM}}$-scheme, described in Sec.3.4. We remind  
that in the class of gauge-invariant MS-like schemes at the $\mathcal{O}(a_s^4)$ level of PT QCD the   GCR relation is defined as  
\begin{equation}
\nonumber
D^{NS}(a_s)C_{Bjp}^{NS}(a_s)=
1+\bigg(\frac{\beta(a_s)}{a_s}\bigg)K(a_s) ~.
\end{equation} 
As already mentioned above in Sec.3.4, the direct computations, performed by us in the $\rm{mMOM}$-scheme with  using Eqs.(\ref{bK2}-\ref{bK3}), lead to the conclusion that in the GCR the factorization of the RG $\beta$-function in the conformal symmetry breaking term is possible only for the  certain specific values of the gauge parameter. 
To study this property in more detail we 
write down the following equality, which allows us to find out what values of the gauge parameter respect the $\beta$-function factorization  property at definite orders of PT:
\begin{equation}
\label{equation}
\frac{\beta^{{\rm{\overline{MS}}}}(a^{{\rm{\overline{MS}}}}_s( a^{{\rm{mMOM}}}_s ))}{a^{{\rm{\overline{MS}}}}_s(a^{{\rm{mMOM}}}_s )}K^{{\rm{\overline{MS}}}}(a^{{\rm{\overline{MS}}}}_s(a^{{\rm{mMOM}}}_s ))=\frac{\beta^{{\rm{mMOM}}}(a^{{\rm{mMOM}}}_s)}{a^{{\rm{mMOM}}}_s}K^{{\rm{mMOM}}}(a^{{\rm{mMOM}}}_s)~.
\end{equation}
This equation  permits us to obtain
the relations between coefficients $K_i$, determined in  the  $\rm{mMOM}$-scheme, and their analogs, evaluated in the $\rm{\overline{MS}}$-scheme. Naturally, the equality (\ref{equation}) is valid only under the assumption of the $\beta$-factorization of the CSB term $\Delta_{csb}$ in Eq.(\ref{BK}) and is not satisfied for any values of $\xi$. Indeed, the Eq.(\ref{equation}) allow us to find all these values of $\xi$, for which the factorization is possible.

Now we find out the criteria for $\beta$-factorization in gauge-dependent $\rm{mMOM}$ scheme. 
Using the expansion $a^{{\rm{\overline{MS}}}}_s$ through  $a^{{\rm{mMOM}}}_s$ in
the  arbitrary gauge (\ref{aMS-mMOM}),  Eq.(\ref{b0Lg}) and  the  relation (\ref{equation}) we reproduce the result, presented above  in Eq.(\ref{K1mMOM}) for any value of $\xi$:
\begin{eqnarray}
\label{K1mMOM-MS}
K^{{\rm{mMOM}}}_1=K^{{\rm{\overline{MS}}}}_1~.
\end{eqnarray}
For the  coefficient $K^{{\rm{mMOM}}}_2$ the relation with $K^{{\rm{\overline{MS}}}}_2$ can be found from Eqs.(\ref{aMS-mMOM}), (\ref{b0Lg}), (\ref{b1Lg}), (\ref{equation})  and looks like as:
\begin{eqnarray}
\label{K2mMOM-MS}
K^{{\rm{mMOM}}}_2=K^{{\rm{\overline{MS}}}}_2+\bigg(\frac{\beta^{{\rm{\overline{MS}}}}_1-\beta^{{\rm{mMOM}}}_1}{\beta_0}+2b^{{\rm{mMOM}}} _1\bigg)K^{{\rm{\overline{MS}}}}_1~.
\end{eqnarray}
In this equation the coefficient 
 $K^{{\rm{mMOM}}}_2$ does not contain terms proportional to $1/\beta_0$,  
if the difference $\beta^{{\rm{\overline{MS}}}}_1-\beta^{{\rm{mMOM}}}_1$  
is proportional to 
the leading coefficient $\beta_0$ of the RG $\beta$-function, namely $\beta^{{\rm{\overline{MS}}}}_1-\beta^{{\rm{mMOM}}}_1=\theta\beta_0C_A$, 
 where $\theta$ is some real number.  Using the explicit expressions for the one and two-loop coefficients of the RG $\beta$-function in 
the $\rm{\overline{MS}}$ and $\rm{mMOM}$ schemes,  we  get the following equations
\begin{eqnarray}
\nonumber
\beta^{{\rm{\overline{MS}}}}_1-\beta^{{\rm{mMOM}}}_1=\bigg(\frac{13}{192}\xi+\frac{5}{96}\xi^2-\frac{1}{64}\xi^3\bigg)C^2_A-\frac{1}{24}(\xi+\xi^2)C_AT_Fn_f=\frac{11}{12}\theta C^2_A-\frac{1}{3}\theta C_AT_Fn_f~.
\end{eqnarray}
Hence, we obtain the following system of equations:
\begin{eqnarray*}
\left\{
\begin{aligned}
\frac{11}{12}\theta &=\frac{13}{192}\xi+\frac{5}{96}\xi^2-\frac{1}{64}\xi^3~,\\
\frac{1}{3}\theta &=\frac{1}{24}\xi+\frac{1}{24}\xi^2~,\\
\end{aligned}
\right.
\end{eqnarray*}
This system leads to the presented above  single equation   (\ref{EQxi}) and has the following solutions
$(\xi, \theta)=(0, 0),$ $(-1, 0),$ $(-3, 3/4)$. Thus, we 
 demonstrate that   
the $\beta$-factorization property  for $\mathcal{O}(a_s^3)$ approximation of the CSB contribution $\Delta_{csb}$ to the GCR  in the $\rm{mMOM}$-scheme  
is possible only for three values of the gauge parameter $\xi$, namely for $\xi=0, -1, -3$\footnote{In the case of $\xi=0,-1$ this conclusion  also follows 
from the fact that for these two values of the gauge parameter  
the two-loop coefficient $\beta^{{\rm{mMOM}}}_1$ coincides with $\beta^{{\rm{\overline{MS}}}}_1$ identically and therefore the 
difference  $\beta^{{\rm{\overline{MS}}}}_1-\beta^{{\rm{mMOM}}}_1$ in Eq.(\ref{K2mMOM-MS}) is  nullified.}. The corresponding $SU(N_c)$ analytical expressions of the 
$K^{{\rm{mMOM}}}_2$ coefficients, defined in Eq.(\ref{function}) in the $\rm{mMOM}$-scheme, have the following form:
\begin{align}
\hspace{-0.5cm}
 \label{xi=0}
K^{{\rm{mMOM}}}_{2, \; \xi=0}&=\bigg[\frac{397}{96}+\frac{17}{2}\zeta_3-15\zeta_5\bigg]C^2_F+\bigg[-\frac{2591}{192}+\frac{91}{8}\zeta_3\bigg]C_FC_A+\bigg[\frac{31}{8}-3\zeta_3\bigg]C_FT_Fn_f \\
\hspace{-0.5cm}
\label{xi=-1}
K^{{\rm{mMOM}}}_{2, \; \xi=-1}&=\bigg[\frac{397}{96}+\frac{17}{2}\zeta_3-15\zeta_5\bigg]C^2_F+\bigg[-\frac{1327}{96}+\frac{47}{4}\zeta_3\bigg]C_FC_A+\bigg[\frac{31}{8}-3\zeta_3\bigg]C_FT_Fn_f \\ 
\hspace{-0.5cm}
\label{xi=-3}
K^{{\rm{mMOM}}}_{2, \; \xi=-3}&=\bigg[\frac{397}{96}+\frac{17}{2}\zeta_3-15\zeta_5\bigg]C^2_F+\bigg[-\frac{695}{48}+\frac{25}{2}\zeta_3\bigg]C_FC_A+\bigg[\frac{31}{8}-3\zeta_3\bigg]C_FT_Fn_f
\end{align}
The  coincidence of the analytical terms, proportional to $C^2_F$-factor, in Eqs.(\ref{xi=0}-\ref{xi=-3}) in the QED-limit with the  results, 
obtained in three  gauge-invariant renormalizations schemes in QED in Eq.(\ref{K2QED}) is quite understandable.  Other typical features of  Eqs.(\ref{xi=0}-\ref{xi=-3}) are  the gauge-dependence of the $C_FC_A$-term to the $K_2^{\rm{mMOM}}$-expression  and the gauge-independence of the $C_FT_Fn_f$-structure. However, as one can see these $C_FT_Fn_f$-contributions differ  from its QED analogs, included in Eq.(\ref{K2QED}). The reason for this difference becomes clear upon careful consideration of the Eq.(\ref{K2mMOM-MS}). Indeed, due to the scheme-independence of the first two coefficients of the RG $\beta$-function in the QED-limit the contribution,  proportional to $\beta^{{\rm{\overline{MS}}}}_1-\beta^{X}_1$, is nullified (here under the $X$ we mean either $\rm{MOM}$ or $\rm{OS}$ renormalization schemes, defined in QED). Unlike the QCD case the second term, proportional to $b^{{\rm{mMOM}}}_1$ (and as a consequence to $n_f$), can always be chosen to be zero in QED due to the corresponding normalizations. That is why these $C_FT_Fn_f$-contributions are 
different.

Considering now the basic transformation (\ref{equation}) from $\rm{\overline{MS}}$ to $\rm{mMOM}$-scheme in the next order of PT we arrive to the following relation between 
$K^{{\rm{mMOM}}}_3$ and  $K^{{\rm{\overline{MS}}}}_3$ coefficients:
\begin{align}
\label{K3mMOM-MS}
&K^{{\rm{mMOM}}}_3=K^{{\rm{\overline{MS}}}}_3+\bigg(\frac{\beta^{{\rm{\overline{MS}}}}_1-\beta^{{\rm{mMOM}}}_1}{\beta_0}+3b^{{\rm{mMOM}}} _1\bigg)K^{{\rm{\overline{MS}}}}_2+ \bigg(2b^{{\rm{mMOM}}} _2+(b^{{\rm{mMOM}}}_1) ^2  \\ \nonumber
&+\frac{\beta^{{\rm{\overline{MS}}}}_2-\beta^{{\rm{mMOM}}}_2}{\beta_0}+\frac{(3\beta^{{\rm{\overline{MS}}}}_1 -2\beta^{{\rm{mMOM}}}_1)b^{{\rm{mMOM}}} _1}{\beta_0}+\frac{ \beta^{{\rm{mMOM}}}_1 (\beta^{{\rm{mMOM}}}_1-\beta^{{\rm{\overline{MS}}}}_1)}{\beta^2_0}
\bigg)K^{{\rm{\overline{MS}}}}_1
\end{align}
Comparing it with Eq.(\ref{K2mMOM-MS}) we notice that  
this relation contains  three  new additional fractions
$(\beta^{{\rm{\overline{MS}}}}_2-\beta^{{\rm{mMOM}}}_2)/\beta_0$, $( 3\beta^{{\rm{\overline{MS}}}}_1 -2\beta^{{\rm{mMOM}}}_1)b^{{\rm{mMOM}}} _1/\beta_0$ and $(\beta^{{\rm{mMOM}}}_1 (\beta^{{\rm{mMOM}}}_1-\beta^{{\rm{\overline{MS}}}}_1))/\beta^2_0$, 
 which did not enter into  Eq.(\ref{K2mMOM-MS}).
 Thus, it is  clear that the question of factorization of the three-loop $\rm{mMOM}$ $\beta$-function in the $\mathcal{O}(a^4_s)$-approximation of the GCR 
reduces to investigating the divisibility of these fractions.
Indeed, if all these fractions are contractible in sum or separately, then we can confidently state the existence of the $\beta$-factorization property in the GCR at $\mathcal{O}(a^4_s)$ level. 
We study this problem for $\xi=0,\; -1,\; -3$ only, since
at these values of gauge parameter 
the $\beta$-factorization property holds at $\mathcal{O}(a^3_s)$ level at least. The extra appearing fraction $(\beta^{{\rm{mMOM}}}_1 (\beta^{{\rm{mMOM}}}_1-\beta^{{\rm{\overline{MS}}}}_1))/\beta^2_0$ is equal to zero in case of $\xi=0$ and $\xi=-1$, and differs from zero for Stefanis--Mikhailov gauge $\xi=-3$. Moreover, at $\xi=-3$ this fraction is irreducible. The remaining two fractions  can not be divided individually at all considered values of $\xi$. However, in Landau gauge the sum of these two fractions are divided by $\beta_0$. Nothing like this is observed for $\xi=-1$ and $\xi=-3$. Therefore we find that the $\beta$-factorization property holds at $\mathcal{O}(a^4_s)$ level in the $\rm{mMOM}$-scheme in Landau gauge $\xi=0$ only and it is not performed for the anti-Feynman gauge $\xi=-1$ and Stefanis--Mikhailov gauge $\xi=-3$. 

Summarizing the foregoing, we get the following analytical expression for $K_3$-term, computed in the Landau gauge:
\begin{align} 
\label{K3mMOM=0}
&K^{{\rm{mMOM}}}_{3, \; \xi=0}=\bigg(\frac{2471}{768}+\frac{61}{8}\zeta_3-\frac{715}{8}\zeta_5+\frac{315}{4}\zeta_7\bigg)C^3_F \\ \nonumber
&+\bigg(\frac{132421}{4608}+\frac{451}{8}\zeta_3-\frac{3685}{48}\zeta_5-\frac{105}{8}\zeta_7\bigg)C^2_FC_A \\ \nonumber
&+\bigg(-\frac{1840145}{18432}+\frac{152329}{3072}\zeta_3+\frac{2975}{48}\zeta_5-\frac{2113}{128}\zeta^2_3\bigg)C_FC^2_A \\ \nonumber
&+\bigg(\frac{71251}{1152}-\frac{539}{24}\zeta_3-\frac{125}{3}\zeta_5+\frac{5}{2}\zeta^2_3\bigg)C_FC_AT_Fn_f \\ \nonumber
&+ \bigg(-\frac{1273}{144}-\frac{599}{24}\zeta_3+\frac{75}{2}\zeta_5\bigg)C^2_FT_Fn_f      +\bigg(-\frac{49}{6}+\frac{7}{2}\zeta_3+5\zeta_5\bigg)C_FT^2_Fn^2_f~.
\end{align} 
For the gauges $\xi=-1$ and $\xi=-3$  the $\rm{mMOM}$
$\mathcal{O}(a_s^4)$ contributions to the  GCR obey the property of the partial $\beta$-factorization only.
It is violated by  the extra  non-factorized contributions, proportional  to the $SU(N_c)$  monomials $C_FC_AT_Fn_f$  in the corresponding 
expressions for the coefficients $K^{{\rm{mMOM}}}_{3, \; \xi=-1}$ and  $K^{{\rm{mMOM}}}_{3, \; \xi=-3}$  (the more  detailed clarification and derivation  of these statements are given in the Appendix B below).  

One should emphasize that relations (\ref{K1mMOM-MS}), (\ref{K2mMOM-MS}), (\ref{K3mMOM-MS}) are valid not only 
for the $\rm{mMOM}$-scheme, but and for any $\rm{MOM}$-like renormalization schemes ($AS$) in QCD. To achieve  this goal  it is only necessary to replace all quantities, calculated in $\rm{mMOM}$ scheme,  by the corresponding quantities in any other $\rm{MOM}$-like scheme 
$(\beta^{\rm{mMOM}}_i, b^{{\rm{mMOM}}} _i)\rightarrow (\beta^{AS}_i, b^{AS}_i)$.

Note, for the QED case all coefficients of $\beta$-function in any renormalization schemes are proportional to the number of charged leptons $N$. 
 Therefore, the  differences $\beta^{{\rm{\overline{MS}}}}_1-\beta^{AS}_1$, $\beta^{{\rm{\overline{MS}}}}_2-\beta^{AS}_2$, $3\beta^{{\rm{\overline{MS}}}}_1-2\beta^{AS}_1$  are always  divided by 
$\beta_0$-factor, and the expression  $\beta^{AS}_1(\beta^{AS}_1-\beta^{{\rm{\overline{MS}}}}_1)$   is always divided by $\beta^2_0$. 
This means that observed  in Sec.2.4   $\beta$-factorization  property of the QED version of the GCR will be valid in  all renormalization schemes in QED.

Moreover, in the case of QED, the  relations (\ref{K2mMOM-MS}), (\ref{K3mMOM-MS}) can be rewritten in  more compact form. Actually, 
taking into account the scheme-independence of the first two coefficients of the RG $\beta$-function of  QED,
the discussed above possibility of fixing the values of   scale parameters, which are leading to nullification  the first  coefficient $b_1$ in the expression  which relates the  QED coupling constants in the $\rm{\overline{MS}}$-scheme with  arbitrary scheme  $X$ (the QED analog of the QCD relation (\ref{aMS-mMOM})) and the special feature,  that in QED the second and third terms $b_i$ in these expressions can be represented through values of corresponding coefficients of $\beta$-function, namely 
\begin{equation}
b^{X}_{2, \;{{\rm{QED}}}}=\frac{\beta^{{\rm{\overline{MS}}}}_2-\beta^{X}_2}{\beta_0}~, ~~~ b^{X}_{3, \;{{\rm{QED}}}}=\frac{\beta^{{\rm{\overline{MS}}}}_3-\beta^{X}_3}{2\beta_0}~,
\end{equation}
we get that in QED the  coefficients $K^X_2$ and $K^X_3$, calculated in the $X$ renormalization scheme, have the following form:
\begin{equation}
\label{confirmation}
K^X_{2, \; {\rm{QED}}}=K^{{\rm{\overline{MS}}}}_{2, \; {\rm{QED}}}~, ~~~~~ K^X_{3, \; {\rm{QED}}}=K^{{\rm{\overline{MS}}}}_{3, \; {\rm{QED}}}+3K_1\frac{\beta^{{\rm{\overline{MS}}}}_2-\beta^X_2}{\beta_0}~.
\end{equation}
The expressions (\ref{confirmation}) are in full agreement with the results, obtained in Sec.2.4 (see also 
the unpublished work, presented in the talk  of  Ref.\cite{GK2012}).

\subsection{The $\beta$-function factorization of the $SU_c(3)$ QCD GCR in the $\mathcal{O}(a_s^3)$ approximation in the  $\rm{MOMgggg}$-scheme}

It is important to find out  whether there are other $\rm{MOM}$-schemes in QCD, which respect the property of 
the RG $\beta$-function factorization in the GCR for the concrete choice of the gauge parameter. Let us  consider  the $\mathcal{O}(a_s^3)$ 
approximation for the GCR in the $\rm{MOMgggg}$ scheme, defined 
by renormalization of the  quartic gluon vertex through subtractions 
of ultraviolet divergences in the symmetric subtraction point,   
investigated and used in the concrete calculations 
in Refs.\cite{Gracey:2014pba},\cite{Gracey:2014ola}.

Using Eq.(\ref{K2mMOM-MS}) and taking into account the explicit form of 
the RG $\beta$-function in this  $\rm{MOMgggg}$-scheme, 
calculated in Ref.\cite{Gracey:2014ola} at the two-loop approximation 
in terms of powers of number of colors $N_c$,  we are convinced that the 
$\mathcal{O}(a^3_s)$ level $\beta$-function factorization property is also 
valid in this scheme  for the  Landau and Stefanis--Mikhailov gauges 
and is violated  for anti-Feynman gauge. Therefore we conclude that 
the property of the factorization of the QCD   $\beta$-function in the $\mathcal{O}(a_s^3)$ approximation for  the fixed anti-Feynman gauge is the 
peculiarity of the $\rm{mMOM}$-scheme. 

In general the two-loop coefficient  for the $\rm{MOMgggg}$-scheme  
QCD  $\beta$-function 
has more complicated analytical expression than its  $\rm{mMOM}$-scheme  analog. Indeed,  the  two-loop coefficient, 
calculated in the $\rm{MOMgggg}$-scheme,  depends on the fourth 
power  of gauge $\xi^4$
and contains additional transcendental logarithmic and Clausen-function  terms  \cite{Gracey:2014ola}. 
The complicated structure remains even in the case of QCD with $SU_c(3)$ group when the Stefanis--Mikhailov gauge is chosen. Indeed, fixing in the results of Ref.\cite{Gracey:2014ola} $N_c=3$ and $\xi=-3$, we get the following analytical expression:
\begin{align}
\beta^{{\rm{MOMgggg}}}_{1, \;\xi=-3, \;N_c=3}&=\frac{4173}{80}+\frac{38907}{800}\log\bigg(\frac{4}{3}\bigg)-\frac{99}{16}\Phi_1\bigg(\frac{3}{4},~\frac{3}{4}\bigg)-\frac{373923}{25600}\Phi_1\bigg(\frac{9}{16},~\frac{9}{16}\bigg) \\ \nonumber
&+\bigg[-\frac{107}{30}-\frac{1179}{400}\log\bigg(\frac{4}{3}\bigg)+\frac{3}{8}\Phi_1\bigg(\frac{3}{4},~\frac{3}{4}\bigg)+\frac{11331}{12800}\Phi_1\bigg(\frac{9}{16},~\frac{9}{16}\bigg)\bigg]n_f~,
\end{align}
where the contributions $\Phi_1(3/4, 3/4)$, $\Phi_1(9/16, 9/16)$ are 
expressed  through  the Clausen function $\rm{Cl_2(\Theta)}$ 
\cite{Gracey:2014ola} as\footnote{One should note that the  analytical 
three-loop QCD results for $R$-ratio in the $\rm{MOMgggg}$-scheme, 
computed in Ref.\cite{Gracey:2014pba},  were expressed through more familiar functions such as the derivative of the logarithm of the Euler $\Gamma$-function and  the imaginary part of polylogarithm function with argument $\exp(iz)/\sqrt{3}$, but not  through $\Phi_1(x, y)$-functions.}:
\begin{align}
\Phi_1\bigg(\frac{3}{4}, \frac{3}{4}\bigg)&=\sqrt{2}\bigg[2{\rm{Cl}}_2\bigg(2\arccos\bigg(\frac{1}{\sqrt{3}}\bigg)\bigg)+{\rm{Cl}}_2\bigg(2\arccos\bigg(\frac{1}{3}\bigg)\bigg)\bigg]~, \\
\Phi_1\bigg(\frac{9}{16}, \frac{9}{16}\bigg)&=\frac{4}{\sqrt{5}}\bigg[2{\rm{Cl}}_2\bigg(2\arccos\bigg(\frac{2}{3}\bigg)\bigg)+{\rm{Cl}}_2\bigg(2\arccos\bigg(\frac{1}{9}\bigg)\bigg)\bigg]~, \\
&~~~~~~~~~~~~{\rm{Cl}}_2(\Theta)=-\int\limits_{0}^{\Theta}dx \log\bigg| 2 \sin \frac{x}{2}\bigg|~.
\end{align}
The  numerical values of these extra terms 
are  $\Phi_1(3/4, 3/4)\approx 2.832045$ and 
$\Phi_1(9/16, 9/16)\approx 3.403614$.

Initially there was no indication that at $\xi=-3$  the 
factorization of the  $\beta$-function in the GCR will be valid  in this scheme at $\mathcal{O}(a^3_s)$ level. However, considering the $\rm{MOMgggg}$ analog of the r.h.s. of Eq.(\ref{K2mMOM-MS}) in the $SU_c(3)$ case, we obtain the following relation:
\begin{equation}
\label{difMOMgggg}
\frac{\beta^{{\rm{\overline{MS}}}}_1-\beta^{{\rm{MOMgggg}}}_{1,~\xi=-3}}{\beta_0}=-\frac{333}{20}-\frac{3537}{200}\log\bigg(\frac{4}{3}\bigg)+\frac{9}{4}\Phi_1\bigg(\frac{3}{4},~\frac{3}{4}\bigg)+\frac{33993}{6400}\Phi_1\bigg(\frac{9}{16},~\frac{9}{16}\bigg)~. 
\end{equation}
Thus, like in the $\rm{mMOM}$-scheme,  in the $\rm{MOMgggg}$-scheme 
in the Stefanis--Mikhailov gauge  
the difference $(\beta^{{\rm{\overline{MS}}}}_1-\beta^{{\rm{MOMgggg}}}_1)$ 
is divided by the $\beta_0$-factor without any residue. 
As was already discussed above,  this feature is essential for the $\beta$-function factorization in the GCR at the $\mathcal{O}(a^3_s)$ level.
Using now the two-loop results of Ref.\cite{Gracey:2014ola} 
for relating  the  coupling constants  of  the $\rm{MOMgggg}$ 
and $\rm{\overline{MS}}$-schemes, taking into account equations 
from the Appendix A, getting the analytical expression for one-loop coefficient in the $\rm{MOMgggg}$-analog of Eq.(\ref{aMS-mMOM}), namely 
$b^{{\rm{MOMgggg}}}_1$-term,  substituting it and  Eq.(\ref{difMOMgggg})
into the  $\rm{MOMgggg}$-version  of relation (\ref{K2mMOM-MS}), 
we find the following analytical expression of the coefficient 
$K_2$, defined at $\xi=-3$:
\begin{eqnarray}
K^{{\rm{MOMgggg}}}_{2,\;\xi=-3,\; N_c=3}&=&-\frac{9337}{270}+\frac{13769}{400}\log\bigg(\frac{4}{3}\bigg)-\frac{35}{32}\Phi_1\bigg(\frac{3}{4},~\frac{3}{4}\bigg)-\frac{2191}{12800}\Phi_1\bigg(\frac{9}{16},~\frac{9}{16}\bigg) \\ \nonumber
&+&\zeta_3\bigg(\frac{2108}{45}-\frac{1967}{50}\log\bigg(\frac{4}{3}\bigg)+\frac{5}{4}\Phi_1\bigg(\frac{3}{4},~\frac{3}{4}\bigg)+\frac{313}{1600}\Phi_1\bigg(\frac{9}{16},~\frac{9}{16}\bigg)\bigg)-\frac{80}{3}\zeta_5 \\ \nonumber
&+&\bigg[\frac{65}{36}-\frac{49}{24}\log\bigg(\frac{4}{3}\bigg)-\frac{49}{96}\Phi_1\bigg(\frac{3}{4},~\frac{3}{4}\bigg)+\frac{7}{96}\Phi_1\bigg(\frac{9}{16},~\frac{9}{16}\bigg) \\ \nonumber
&+& \zeta_3\bigg(-\frac{10}{9}+\frac{7}{3}\log\bigg(\frac{4}{3}\bigg)+\frac{7}{12}\Phi_1\bigg(\frac{3}{4},~\frac{3}{4}\bigg)+\frac{1}{12}\Phi_1\bigg(\frac{9}{16},~\frac{9}{16}\bigg)\bigg)\bigg]n_f~.
\end{eqnarray}
In the case of the Landau gauge $\xi=0$ all transcendental functions 
included into the  two-loop coefficient of the $\rm{MOMgggg}$ RG 
$\beta$-function are nullified and we arrive to the following result
\begin{equation}
\beta^{{\rm{MOMgggg}}}_{1,\;\xi=0,\;N_c=3}=\frac{51}{8}-\frac{19}{24}n_f~,
\end{equation}
which as expected is equal to the  $\rm{\overline{MS}}$-expression 
for the  $\beta_1$-coefficient. 
Following the described above considerations, we also  
obtain the following expression for $K_2$-coefficient in the Landau gauge:
 \begin{eqnarray}
 \hspace{-0.3cm}
 K^{{\rm{MOMgggg}}}_{2,\;\xi=0,\;N_c=3}&=&-\frac{280073}{8640}+\frac{3017}{100}\log\bigg(\frac{4}{3}\bigg)-\frac{595}{256}\Phi_1\bigg(\frac{3}{4},~\frac{3}{4}\bigg)-\frac{50533}{51200}\Phi_1\bigg(\frac{9}{16},~\frac{9}{16}\bigg) \\ \nonumber
 &+& \zeta_3\bigg(\frac{15973}{360}-\frac{862}{25}\log\bigg(\frac{4}{3}\bigg)+\frac{85}{32}\Phi_1\bigg(\frac{3}{4},~\frac{3}{4}\bigg)+\frac{7219}{6400}\Phi_1\bigg(\frac{9}{16},~\frac{9}{16}\bigg)\bigg)-\frac{80}{3}\zeta_5 \\ \nonumber
 &+&\bigg[\frac{65}{36}-\frac{49}{24}\log\bigg(\frac{4}{3}\bigg)-\frac{49}{96}\Phi_1\bigg(\frac{3}{4},~\frac{3}{4}\bigg)+\frac{7}{96}\Phi_1\bigg(\frac{9}{16},~\frac{9}{16}\bigg) \\ \nonumber
 &+& \zeta_3\bigg(-\frac{10}{9}+\frac{7}{3}\log\bigg(\frac{4}{3}\bigg)+\frac{7}{12}\Phi_1\bigg(\frac{3}{4},~\frac{3}{4}\bigg)+\frac{1}{12}\Phi_1\bigg(\frac{9}{16},~\frac{9}{16}\bigg)\bigg)\bigg]n_f~.
 \end{eqnarray}
Thus the consideration of completely different gauge-dependent renormalization schemes, namely the $\rm{mMOM}$ and $\rm{MOMgggg}$ schemes, allows us to discover that at $\xi=0$ and $\xi=-3$ the factorization of the RG $\beta$-function holds in the GCR at $\mathcal{O}(a^3_s)$ level. 
Therefore,  it is important  to understand why this  happens 
not only for the  Landau  gauge, where the two-loop coefficient of 
the QCD $\beta$-function coincides  with the one  
of the $\rm{\overline{MS}}$-scheme,  but and for the    Stefanis--Mikhailov 
gauge as well. 
Next,  in Sec.4.1 we have shown that in the Landau gauge 
only the $\beta$-function factorization property is valid in the  
$\rm{mMOM}$-scheme at the $\mathcal{O}(a^4_s)$ level. 
These foundations give us the hint that  
for the  gauges $\xi=0$ and $\xi=-3$ the fundamental property of 
factorization will be true  at the $\mathcal{O}(a^3_s)$ level 
in $\textit{all}$ $\rm{MOM}$-like schemes, while for the Landau gauge 
this property will be fulfilled 
in $\textit{all}$ $\rm{MOM}$-like schemes in the $\mathcal{O}(a^4_s)$ 
approximation at least. Moreover, we may make the guess that in the higher orders of PT the property of $\beta$-function factorization will take place in $\rm{MOM}$-like schemes in the case of Landau gauge as well. 
In the next section we will study these assumptions  in more detail. 

\subsection{Special features of the Landau and Stefanis-Mikhailov gauges in the QCD $\rm{MOM}$-like schemes}
The summarized  above assumptions 
can be proved  using the   following relation, 
which is similar to the one of Eq.(\ref{bmMOM-MS}):
\begin{equation}
\label{allschemes}
\beta^{AS}(a^{AS}_s, \xi^{AS})=\beta^{\rm{\overline{MS}}}(a_s^{{\rm{\overline{MS}}}})\frac{\partial a^{AS}_s}{\partial a^{\rm{\overline{MS}}}_s}+\xi^{\rm{\overline{MS}}}\gamma^{\rm{\overline{MS}}}_{\xi}(a_s^{{\rm{\overline{MS}}}}, \xi^{\rm{\overline{MS}}}) \frac{\partial a^{AS}_s}{\partial \xi^{\rm{\overline{MS}}}}\bigg\vert_{{\rm{\overline{MS}}}\rightarrow {AS}}
\end{equation}
where $AS$ denotes any $\rm{MOM}$-like renormalization scheme with 
linear covariant gauge. Considering Eq.(\ref{allschemes})   
and the arbitrary formal PT series, containing the expansion 
of the coupling constant,  defined in the $\rm{\overline{MS}}$-scheme, 
through  the  coupling constant of  another  $AS$-scheme, 
namely $a^{\rm{\overline{MS}}}_s=a^{AS}_s+\sum\limits_{k=1} b_k (a^{AS}_s)^{k+1}$, and taking into account the formulas of Appendix A, we get:
 \begin{equation}
 \label{betaAStoMS1all}
 \beta^{AS}_1=\beta^{\rm{\overline{MS}}}_1-\xi\gamma^{\rm{\overline{MS}}}_0(\xi)\frac{\partial b_1(\xi)}{\partial\xi}~,
 \end{equation}
where $\xi=\xi^{AS}$ and  $\gamma^{\rm{\overline{MS}}}_0=(-13/24+\xi^{\rm{\overline{MS}}}/8)C_A+T_Fn_f/3$  is the one-loop coefficient of the anomalous dimension of 
gauge.  Due to its 
 linear covariance it  coincides with the corresponding coefficient of the 
gluon field anomalous dimension but with the opposite sign. 
As  follows from Sec.4.1, to   study the validity of the 
property of the  $\beta$-function factorization in the GCR 
at the $\mathcal{O}(a_s^3)$-level  we should  consider the  issue  
whether the   expression for  
$(\beta^{\rm{\overline{MS}}}_1-\beta^{AS}_1)$ is divided by  $\beta_0$-factor.
It is clear from Eq.(\ref{betaAStoMS1all}) that at $\xi=0$ $\beta^{AS}=\beta^{\rm{\overline{MS}}}_1$ and therefore the factorization of $\beta$-function in the 
GCR is always possible in the  Landau gauge at this level in the  $AS$ scheme. 
Next, since at $\xi=-3$ we have  $\gamma_0=-11C_A/12+T_Fn_f/3=-\beta_0$, it is obviously from Eq.(\ref{betaAStoMS1all}) that in the Stefanis--Mikhailov gauge the expression $(\beta^{\rm{\overline{MS}}}_1-\beta^{AS}_1)$ will always  be divided by $\beta_0$-factor.\footnote{The fact that at $\xi=-3$ the one-loop expression for renormalization constant of gluon field is proportional to $\beta_0$ was noted in Refs.\cite{Mikhailov:1998xi, Mikhailov:1999ht}.}
Therefore  at $\xi=-3$ the $\beta$-factorization property is always 
valid at the $\mathcal{O}(a^3_s)$ level in the GCR in the  $AS$ scheme 
with linear covariant gauge. Note, that  
these statements do not contradict 
 that the $\beta$-function factorization may also hold 
 at other values of $\xi$. Indeed, as was already demonstrated  
above this feature holds in the $\rm{mMOM}$-scheme at $\xi=-1$ as well. 
Now we can make more strict statement  that other values of $\xi$,  which respect the $\beta$-function   factorization property  at the  $\mathcal{O}(a^3_s)$ level,  depend on the  concrete  renormalization scheme and are determined by the special behavior of the $\partial b_1/\partial \xi$-term.

Taking into account the PT relation between  gauge parameters, defined in the $\rm{\overline{MS}}$ and arbitrary $AS$ $\rm{MOM}$-like scheme, namely
$\xi^{{\rm{\overline{MS}}}}=\xi^{AS}+\xi^{AS}\sum\limits_{k=1}\eta_k(a^{AS}_s)^k$, and using the transformation equations from Appendix A, we obtain the following relation  between three-loop  coefficients of the $AS$  and  $\rm{\overline{MS}}$-schemes: 
\begin{align}
\label{betaAS-MSxi}
\beta^{AS}_2&=\beta^{\rm{\overline{MS}}}_2+\beta^{\rm{\overline{MS}}}_1b_1(\xi)+\beta_0\bigg(b^2_1(\xi)-b_2(\xi)\bigg)-\xi\gamma^{\rm{\overline{MS}}}_0(\xi)\frac{\partial b_2(\xi)}{\partial\xi} \\ \nonumber
&+\xi\frac{\partial b_1(\xi)}{\partial\xi}\bigg(\beta_0\eta_1(\xi)+\gamma^{\rm{\overline{MS}}}_0(\xi)b_1(\xi)-\gamma^{\rm{\overline{MS}}}_1(\xi)+\xi\gamma^{\rm{\overline{MS}}}_0(\xi)\frac{\partial\eta_1(\xi)}{\partial\xi}-\xi\eta_1(\xi)\frac{\partial\gamma^{\rm{\overline{MS}}}_0(\xi)}{\partial\xi}\bigg)~.
\end{align}
Here $\xi=\xi^{AS}$, $\gamma^{\rm{\overline{MS}}}_1(\xi)$ is the two-loop coefficient of the anomalous dimension of  gauge \cite{Chetyrkin:2000dq}. The application of the relation (\ref{betaAS-MSxi}) allows to confirm the results of  calculations of  the three-loop coefficient of $\beta$-function in the $\rm{mMOM}$-scheme \cite{vonSmekal:2009ae},\cite{Gracey:2013sca}. 
Since we have already found out that the factorization of $\beta$-function  
does not occur at the  $\mathcal{O}(a^4_s)$ level in the 
$\rm{mMOM}$-scheme at $\xi=-3$, at this PT level   
the Landau gauge will be  only applied. 
At $\xi=0$ in an  arbitrary $AS$-scheme we obtain the following  simplified version of Eq.(\ref{betaAS-MSxi}):
\begin{equation}
\label{LandaubetaAS}
\beta^{AS}_{2, \; \xi=0}=\beta^{\rm{\overline{MS}}}_2+\beta^{\rm{\overline{MS}}}_1b_1(0)+\beta_0\bigg(b^2_1(0)-b_2(0)\bigg)~.
\end{equation}
In the limit $\xi=0$ the analog of the relation (\ref{K3mMOM-MS}) has the following form:
\begin{align}
\label{K3AS-MS-Landau}
K^{AS}_3=K^{{\rm{\overline{MS}}}}_3+3b_1(0)K^{{\rm{\overline{MS}}}}_2+ \bigg(2b_2(0)+b^2_1(0)+\frac{\beta^{{\rm{\overline{MS}}}}_2-\beta^{AS}_2}{\beta_0}+\frac{\beta^{{\rm{\overline{MS}}}}_1b_1(0) }{\beta_0}
\bigg)K^{{\rm{\overline{MS}}}}_1~.
\end{align}
Substituting the expression for $\beta^{AS}_{2, \; \xi=0}$ from Eq.(\ref{LandaubetaAS}) into Eq.(\ref{K3AS-MS-Landau}), we find that
\begin{align}
\label{K3AS-MS-Landau-xi=0}
K^{AS}_3=K^{{\rm{\overline{MS}}}}_3+3b_1(0)K^{{\rm{\overline{MS}}}}_2+ 3b_2(0)K^{{\rm{\overline{MS}}}}_1~.
\end{align}
Thus, the relation (\ref{K3AS-MS-Landau-xi=0}) proves the validity of the factorization of the RG $\beta$-function in the Landau gauge in the GCR in arbitrary $AS$ $\rm{MOM}$-type scheme at the $\mathcal{O}(a^4_s)$ level.

This fact leads us to the natural  assumption of the realization of the property of  the $\beta$-function factorization in 
the GCR, evaluated in the  $AS$ $\rm{MOM}$-like  scheme with linear covariant gauge  in any order of perturbation theory in the Landau gauge. Indeed, 
using the  relation (\ref{allschemes}), we obtain the following expressions for the four- and five-loop coefficients of $\beta$-function in $AS$-scheme when the Landau gauge is chosen:
\begin{align}
\label{beta3LandauAS}
\beta^{AS}_{3, \; \xi=0}&=\beta^{\rm{\overline{MS}}}_3+2\beta^{\rm{\overline{MS}}}_2b_1(0)+\beta^{\rm{\overline{MS}}}_1b^2_1(0)+2\beta_0\bigg(2b_1(0)b_2(0)-b^3_1(0)-b_3(0)\bigg)~, \\
\label{beta4LandauAS}
\beta^{AS}_{4, \; \xi=0}&=\beta^{\rm{\overline{MS}}}_4+\beta^{\rm{\overline{MS}}}_2\bigg(b_2(0)+2b^2_1(0)\bigg)+\beta^{\rm{\overline{MS}}}_1\bigg(3b_1(0)b_2(0)-b^3_1(0)-b_3(0)\bigg) \\ \nonumber
&+3\beta^{\rm{\overline{MS}}}_3b_1(0)+\beta_0\bigg(4b^4_1(0)-11b_2(0)b^2_1(0)+6b_3(0)b_1(0)+
4b^2_2(0)-3b_4(0)\bigg)~.
\end{align}
Keeping in mind the presented in Refs.\cite{Gabadadze:1995ei,Crewther:1997ux,Braun:2003rp} arguments
in favor of the validity of the  factorization of the RG $\beta$-function
  for the GCR  in gauge-invariant $\rm{\overline{MS}}$-scheme  in all orders of PT, using Eqs.(\ref{K1mMOM-MS}),(\ref{K2mMOM-MS}),(\ref{LandaubetaAS}),(\ref{K3AS-MS-Landau-xi=0}-\ref{beta4LandauAS}) and taking into account the relation (\ref{equation}), we obtain the $\mathcal{O}(a^5_s)$ and $\mathcal{O}(a^6_s)$-coefficients of the  
polynomial $K(a_s)$ of Eq.(\ref{function}), which are
  defined in the  $AS$-scheme at $\xi=0$:
\begin{align}
\label{K4AS}
K^{AS}_4&=K^{{\rm{\overline{MS}}}}_4+4b_1(0)K^{{\rm{\overline{MS}}}}_3+\bigg(4b_2(0)+2b^2_1(0)\bigg)K^{{\rm{\overline{MS}}}}_2+4b_3(0)K^{{\rm{\overline{MS}}}}_1~, \\
\label{K5AS}
K^{AS}_5&=K^{{\rm{\overline{MS}}}}_5+5b_1(0)K^{{\rm{\overline{MS}}}}_4+5\bigg(b_2(0)+b^2_1(0)\bigg)K^{{\rm{\overline{MS}}}}_3+5\bigg(b_3(0)+b_1(0)b_2(0)\bigg)K^{{\rm{\overline{MS}}}}_2 \\ \nonumber
&+5b_4(0)K^{{\rm{\overline{MS}}}}_1~.
\end{align}
Unlike the relations (\ref{K2mMOM-MS}), (\ref{K3mMOM-MS}) the obtained expressions (\ref{K4AS}-\ref{K5AS}) do not contain the terms, proportional to the powers of $1/\beta_0$-factor. Therefore we conclude that in the $\rm{MOM}$-type schemes in the Landau gauge the factorization of $\beta$-function holds in the GCR at the $\mathcal{O}(a^5_s)$ and $\mathcal{O}(a^6_s)$ level. There are no obstacles to obtain similar expressions for any value of order of PT. Thus we conclude  that  since  the  $\beta^{{\rm{\overline{MS}}}}$-function factorization property of GCR is  most probably true in all  orders of PT, this property  also takes 
place in Landau gauge in all  $\rm{MOM}$-type schemes  in any  order of PT.  
The  explanation of  this  new  feature    
from the first principles of perturbative QCD 
is the important   opened problem.

\section{Conclusion}

In this work we find out that the property of factorization of the RG $\beta$-function in the generalized Crewther relation really holds true 
not only in gauge-invariant renormalization schemes in QCD and QED, but  in the gauge-dependent subtraction schemes in QCD as well.
Considering two gauge non-invariant QCD schemes, namely  $\rm{mMOM}$- and  $\rm{MOMgggg}$-schemes, we discover that the $\beta$-factorization 
property of the GCR is valid at the  $\mathcal{O}(a^3_s)$ level in all $\rm{MOM}$-like 
renormalization schemes with linear covariant gauge at the Landau $\xi=0$ and the Stefanis--Mikhailov  $\xi=-3$ gauges. 
We also find out that in the $\rm{mMOM}$-scheme  in this order of PT approximation the property of the $\beta$-function factorization is also valid in 
one more extra gauge, namely anti-Feynman gauge with $\xi=-1$. However, in the $\mathcal{O}(a^4_s)$ approximation the $\rm{mMOM}$ $SU(N_c)$ 
expressions for the GCR with $\xi=-3$ and $\xi=-1$ satisfy the property of the partial $\beta$-function factorization only, whereas in Landau gauge this 
problem do not manifest itself. Moreover, we conclude that if the factorization of the RG $\beta$-function in the GCR in the $\rm{\overline{MS}}$-scheme 
persists in all orders of PT, then it is also true  in all $\rm{MOM}$-like scheme with linear covariant Landau  gauge.
 Therefore the gauge-invariance of the renormalization schemes is $\textit{sufficient  but not necessary}$
condition for the manifestation of the $\beta$-function factorization in the conformal symmetry breaking term of the QCD  GCR. 
We obtain the explicit $SU(N_c)$ analytical $\mathcal{O}(a^4_s)$ approximations  for Adler and Bjorken functions in the $\rm{mMOM}$-scheme
with the arbitrary covariant gauge.  We also  show that in QED  the $\beta$-factorization property remains valid in all orders of PT 
in any ultraviolet subtraction scheme.  We expect, that the similar QCD and QED properties will be true for the proposed in Ref.\cite{Kataev:2010du} 
variant of the GCR, with its conformal symmetry breaking term, 
expressed through the two-fold series in powers of the conformal anomaly and the coupling constant of these fundamental gauge theories.

\acknowledgments
The authors are grateful to G.Cveti\v c,  S.V.Mikhailov and N.G.Stefanis for useful discussions.  
The work of A.K and V.M  is  
supported by  the Russian Science Foundation Grant No. 14-22-00161.

\appendix
\section{The transformation  relations}
In this part we present transformation relations, which 
 allow us to obtain the PT coefficients  in one particular renormalization scheme, if coefficients in another scheme are known. Let us explain this in details, considering 
 the following example. Suppose that we know the explicit  form of the coefficients $\upsilon_i(\overline{\xi})$ and $\omega_i(\overline{\xi})$, depending on gauge parameter $\overline{\xi}$ defined in some bar-renormalization scheme  $\rm{\overline{AS}}$, in the  following expressions:
\begin{eqnarray}
a_s&=&\overline{a}_s\bigg(1+\upsilon_1(\overline{\xi})\overline{a}_s+\upsilon_2(\overline{\xi})\overline{a}^2_s+\upsilon_3(\overline{\xi})\overline{a}^3_s+\mathcal{O}(\overline{a}^4_s)\bigg)~, \\
\xi &=&\overline{\xi}\bigg(1+\omega_1(\overline{\xi})\overline{a}_s+\omega_2(\overline{\xi})\overline{a}^2_s+\mathcal{O}(\overline{a}^3_s)\bigg)~.
\end{eqnarray}
Our aim is to find analogies $b_i(\xi)$, $\eta_i(\xi)$ of  coefficients $\upsilon_i(\overline{\xi})$ and $\omega_i(\overline{\xi})$ correspondingly, determined in another scheme $\rm{AS}$ at the same order of perturbation theory:
\begin{eqnarray}
\overline{a}_s&=&a_s\bigg(1+b_1(\xi)a_s+b_2(\xi)a^2_s+b_3(\xi)a^3_s+\mathcal{O}(a^4_s)\bigg)~, \\
\overline{\xi}&=&\xi\bigg(1+\eta_1(\xi)a_s+\eta_2(\xi)a^2_s+\mathcal{O}(a^3_s)\bigg)~.
\end{eqnarray}
Using the Taylor expansion
it is straightforward to obtain the following set of transformation equations:
\begin{eqnarray}
\label{v1}
b_1(\xi)&=&-\upsilon_1(\xi)~, ~~~ \eta_1(\xi)=-\omega_1(\xi)~, \\
\label{v2}
b_2(\xi)&=&-\upsilon_2(\xi)+2\upsilon^2_1(\xi)+\xi\omega_1(\xi)\frac{d\upsilon_1(\xi)}{d\xi}~, \\
\label{v3}
\eta_2(\xi)&=&-\omega_2(\xi)+\omega^2_1(\xi)+\omega_1(\xi)\upsilon_1(\xi)+\xi\omega_1(\xi)\frac{d\omega_1(\xi)}{d\xi}~, \\
\label{v4}
b_3(\xi)&=&-\upsilon_3(\xi)+5\upsilon_2(\xi)\upsilon_1(\xi)-5\upsilon^3_1(\xi)+\xi\omega_1(\xi)\frac{d\upsilon_2(\xi)}{d\xi}-\frac{1}{2}\xi^2\omega^2_1(\xi)\frac{d^2\upsilon_1(\xi)}{d\xi^2} \\ \nonumber
&+&\xi\frac{d\upsilon_1(\xi)}{d\xi}\bigg(\omega_2(\xi)-\omega^2_1(\xi)-5\omega_1(\xi)\upsilon_1(\xi)-\xi\omega_1(\xi)\frac{d\omega_1(\xi)}{d\xi}\bigg)~.
\end{eqnarray}
The presented relations (\ref{v1}--\ref{v4}) solve  the stated problem of finding the required coefficients $b_i(\xi)$ and $\eta_i(\xi)$ in the $\rm{AS}$-scheme (note, that in the context of our studies, presented in this work, under the $\rm{AS}$ scheme  we mean either the $\rm{mMOM}$ or $\rm{MOMgggg}$-schemes, and  under the $\rm{\overline{AS}}$ we imply the $\rm{\overline{MS}}$-scheme).

\section{About partial factorization  in the $\rm{mMOM}$-scheme at  the $\mathcal{O}(a^4_s)$ level}
\subsection{The case of the Stefanis--Mikhailov gauge $\xi=-3$}
  
Using the  relations (\ref{bK3}), (\ref{b1Lg}), (\ref{b2Lg}), (\ref{DmMOMadl}), (\ref{CmMOMBjr}), (\ref{K1mMOM}), (\ref{xi=-3})
 we obtain the following $\rm{mMOM}$-expression, defined at the $\mathcal{O}(a^4_s)$ 
level in the Stefanis--Mikhailov gauge:
\begin{gather}
\label{App3}
d_4+c_4+d_1c_3+c_1d_3+d_2c_2+\beta_1K_2+\beta_2K_1\bigg\vert^{{\rm{mMOM}}}_{\xi=-3} = \\ \nonumber
=\bigg[-\frac{27181}{9216}-\frac{671}{96}\zeta_3+\frac{7865}{96}\zeta_5-\frac{1155}{16}\zeta_7\bigg]C^3_FC_A+\bigg[\frac{2471}{2304}+\frac{61}{24}\zeta_3-\frac{715}{24}\zeta_5+\frac{105}{4}\zeta_7\bigg]C^3_FT_Fn_f \\ \nonumber
+\bigg[-\frac{747967}{27648}-\frac{20405}{384}\zeta_3+\frac{10505}{144}\zeta_5+\frac{385}{32}\zeta_7\bigg]C^2_FC^2_A +\bigg[-\frac{1273}{432}-\frac{599}{72}\zeta_3+\frac{25}{2}\zeta_5\bigg]C^2_FT^2_Fn^2_f \\ \nonumber
+\bigg[\frac{10031531}{110592}-\frac{376445}{9216}\zeta_3-\frac{32725}{576}\zeta_5+\frac{4499}{384}\zeta^2_3\bigg]C_FC^3_A+\bigg[-\frac{49}{18}+\frac{7}{6}\zeta_3+\frac{5}{3}\zeta_5\bigg]C_FT^3_Fn^3_f \\ \nonumber
+\bigg[\frac{124009}{6912}+\frac{6077}{144}\zeta_3-\frac{4385}{72}\zeta_5-\frac{35}{8}\zeta_7\bigg]C^2_FC_AT_Fn_f\\ \nonumber
+\bigg[-\frac{2499749}{27648}+\frac{27673}{768}\zeta_3+\frac{2825}{48}\zeta_5-\frac{629}{96}\zeta^2_3\bigg]C_FC^2_AT_Fn_f \\ \nonumber
+\bigg[\frac{12245}{432}-\frac{1567}{144}\zeta_3-\frac{665}{36}\zeta_5+\frac{5}{6}\zeta^2_3\bigg]C_FC_AT^2_Fn^2_f~.
\end{gather}
According to Eq.(\ref{bK3}) in the case of existence of the 
 $\beta$-function factorization property at the  $\mathcal{O}(a^4_s)$ level of the GCR the expression (\ref{App3}) must be equal to the $-\beta_0K_3$ term at $\xi=-3$. In this case 
 the coefficient  $K_3$  must contain the same group structures as in the Landau gauge (\ref{K3mMOM=0}), namely
\begin{equation}
\label{structure}
K_3=\theta_1C^3_F+\theta_2C^2_FC_A+\theta_3C_FC^2_A+\theta_4C^2_FT_Fn_f+\theta_5C_FC_AT_Fn_f+\theta_6C_FT^2_Fn^2_f~,
\end{equation}
where $\theta_1-\theta_6$ are unknown analytical coefficients.
Taking into account expression (\ref{b0Lg}) one can find the following expression:
\begin{eqnarray*}
-\beta_0K_3&=&-\frac{11}{12}\theta_1C^3_FC_A-\frac{11}{12}\theta_2C^2_FC^2_A-\frac{11}{12}\theta_3C_FC^3_A+\frac{1}{3}\theta_1C^3_FT_Fn_f \\
&+&\bigg(\frac{1}{3}\theta_2-\frac{11}{12}\theta_4\bigg)C^2_FC_AT_Fn_f 
+\bigg(\frac{1}{3}\theta_3-\frac{11}{12}\theta_5\bigg)C_FC^2_AT_Fn_f \\
&+&\bigg(\frac{1}{3}\theta_5-\frac{11}{12}\theta_6\bigg)C_FC_AT^2_Fn^2_f 
+\frac{1}{3}\theta_4C^2_FT^2_Fn^2_f+\frac{1}{3}\theta_6C_FT^3_Fn^3_f
\end{eqnarray*}
Equating the expression (\ref{App3}) with $-\beta_0K_3$ term we obtain for Stefanis--Mikhailov gauge:
\begin{eqnarray*}
\theta^{\xi=-3}_1&=&\frac{2471}{768}+\frac{61}{8}\zeta_3-\frac{715}{8}\zeta_5+\frac{315}{4}\zeta_7~, \\
\theta^{\xi=-3}_2&=&\frac{67997}{2304}+\frac{1855}{32}\zeta_3-\frac{955}{12}\zeta_5-\frac{105}{8}\zeta_7~, \\
\theta^{\xi=-3}_3&=&-\frac{10031531}{101376}+\frac{376445}{8448}\zeta_3+\frac{2975}{48}\zeta_5-\frac{409}{32}\zeta^2_3~, \\
\theta^{\xi=-3}_4&=&-\frac{1273}{144}-\frac{599}{24}\zeta_3+\frac{75}{2}\zeta_5~, \\
\theta^{\xi=-3}_6&=&-\frac{49}{6}+\frac{7}{2}\zeta_3+5\zeta_5~.
\end{eqnarray*}
The system of equations containing the $\theta_5$  contribution at the group weights $C_FC^2_AT_Fn_f$ and $C_FC_AT^2_Fn^2_f$ in $-\beta_0K_3$ term is incompatible. Therefore
the coefficient $\theta_5$ can not be found indicating the absence of the $\beta$ factorization for $\xi=-3$ at $\mathcal{O}(a^4_s)$ level. However, one should emphasize that there holds a partial factorization for five of the six possible coefficients $\theta_i$ in group structures in (\ref{structure}) except the $\theta_5C_FC_AT_Fn_f$ term.
 
\subsection{The case of the anti-Feynman gauge $\xi=-1$}
Similarly one can obtain:
\begin{gather}
\label{App1}
d_4+c_4+d_1c_3+c_1d_3+d_2c_2+\beta_1K_2+\beta_2K_1\bigg\vert^{{\rm{mMOM}}}_{\xi=-1} = \\ \nonumber
=\bigg[-\frac{27181}{9216}-\frac{671}{96}\zeta_3+\frac{7865}{96}\zeta_5-\frac{1155}{16}\zeta_7\bigg]C^3_FC_A+\bigg[\frac{2471}{2304}+\frac{61}{24}\zeta_3-\frac{715}{24}\zeta_5+\frac{105}{4}\zeta_7\bigg]C^3_FT_Fn_f \\ \nonumber
+\bigg[-\frac{747967}{27648}-\frac{20405}{384}\zeta_3+\frac{10505}{144}\zeta_5+\frac{385}{32}\zeta_7\bigg]C^2_FC^2_A +\bigg[-\frac{1273}{432}-\frac{599}{72}\zeta_3+\frac{25}{2}\zeta_5\bigg]C^2_FT^2_Fn^2_f \\ \nonumber
+\bigg[\frac{1287481}{13824}-\frac{26255}{576}\zeta_3-\frac{32725}{576}\zeta_5+\frac{335}{24}\zeta^2_3\bigg]C_FC^3_A+\bigg[-\frac{49}{18}+\frac{7}{6}\zeta_3+\frac{5}{3}\zeta_5\bigg]C_FT^3_Fn^3_f \\ \nonumber
+\bigg[\frac{124009}{6912}+\frac{6077}{144}\zeta_3-\frac{4385}{72}\zeta_5-\frac{35}{8}\zeta_7\bigg]C^2_FC_AT_Fn_f \\ \nonumber
+\bigg[-\frac{2523437}{27648}+\frac{9671}{256}\zeta_3+\frac{2825}{48}\zeta_5-\frac{713}{96}\zeta^2_3\bigg]C_FC^2_AT_Fn_f \\ \nonumber
+\bigg[\frac{12245}{432}-\frac{1567}{144}\zeta_3-\frac{665}{36}\zeta_5+\frac{5}{6}\zeta^2_3\bigg]C_FC_AT^2_Fn^2_f~.
\end{gather}
It is interesting to note that in this expression 
among the nine possible terms in $SU(N_c)$ group structures the seven coincide with the results for Stefanis--Mikhailov gauge, namely contributions which is proportional to $C^3_FC_A,$ $C^2_FC^2_A,$ $C^3_FT_Fn_f,$  $C^2_FC_AT_Fn_f,$ $C_FC_AT^2_Fn^2_f,$  $C^2_FT^2_Fn^2_f,$  $C_FT^3_Fn^3_f$ in anti-Feynman gauge are identically equal to the results for $\xi=-3$ presented in (\ref{App3}).
Separately it should be explained that of these seven contributions, the three are gauge--dependent, namely  those that are proportional to 
 $C^2_FC^2_A,$ $C^2_FC_AT_Fn_f,$ $C_FC_AT^2_Fn^2_f$ group weights. Nevertheless it is quite surprising that these gauge-dependent terms coincide at $\xi=-3$ and $\xi=-1$.

For $\xi=-1$ we obtain the following values of the $\theta_i$ coefficients:
\begin{eqnarray*}
\theta^{\xi=-1}_1&=&\theta^{\xi=-3}_1 ~, ~~~~~
\theta^{\xi=-1}_2=\theta^{\xi=-3}_2~, ~~~~~ \theta^{\xi=-1}_4=\theta^{\xi=-3}_4~, ~~~~~ \theta^{\xi=-1}_6=\theta^{\xi=-3}_6~, \\
\theta^{\xi=-1}_3&=&-\frac{1287481}{12672}+\frac{26255}{528}\zeta_3+\frac{2975}{48}\zeta_5-\frac{335}{22}\zeta^2_3~.
\end{eqnarray*}
The $\beta$ factorization property is again violated by the term $\theta_5C_FC_AT_Fn_f$ in expansion (\ref{structure}). Therefore in this case we observe the partial factorization of the $\mathcal{O}(a^4_s)$ $\rm{mMOM}$-scheme approximation of the GCR as well.

\end{document}